\documentclass[twoside,twocolumn,english,prc,amsmath,a4paper,myheadings]{article}
\usepackage{mathpazo}
\usepackage[LGR,T1]{fontenc}
\usepackage[latin9]{inputenc}
\usepackage{geometry}
\usepackage{verbatim}
\usepackage{cprotect}
\usepackage{float}
\usepackage{amsmath}
\usepackage{amssymb}
\usepackage{graphicx}

\makeatletter

\DeclareRobustCommand{\greektext}{%
  \fontencoding{LGR}\selectfont\def\encodingdefault{LGR}}
\DeclareRobustCommand{\textgreek}[1]{\leavevmode{\greektext #1}}

\newcommand{\lyxaddress}[1]{
	\par {\raggedright #1
	\vspace{1.4em}
	\noindent\par}
}

\def\@oddhead{\rightmark \hfill Flow in small systems in EPOS4 \hfill \thepage}
\def\@evenhead{\thepage \hfill Klaus WERNER\hfill}
\def\fnum@table{\tablename~{\bf\thetable}}
\def\fnum@figure{\figurename~{\bf\thefigure}}
\def\tablename{\footnotesize{\bf Table}}
\def\figurename{\footnotesize{\bf Figure}}

\usepackage{dcolumn}
\usepackage[font=small,labelfont=bf]{caption}

\def\citet{\cite}

\AtBeginDocument{
  
}

\makeatother

\usepackage{babel}
\begin{document}
\twocolumn[   \begin{@twocolumnfalse}  
\title{Flow in small systems in the EPOS4 approach for high-energy scatterings}
\author{{\normalsize K.$\,$Werner}}
\date{}
\maketitle

\lyxaddress{\begin{center}
SUBATECH, Nantes University \textendash{} IN2P3/CNRS \textendash{}
IMT Atlantique, Nantes, France
\par\end{center}}
\begin{abstract}
EPOS4 is based on a sophisticated (recently significantly improved) parallel-primary-scattering scenario followed by a hydrodynamic expansion, for all collision systems, from small ones such as proton-proton ($pp$) to big ones such as lead-lead (PbPb). Having already reported on identified particle spectra in recent publications (providing information about radial flow), I discuss here the multiplicity dependence of multi-particle cumulants and flow harmonics, to better understand collectivity in small systems. The model is not particularly tuned for flow results, but it is a "general purpose" approach, trying to accommodate various types of observables with the same model.

~~
\end{abstract}
\end{@twocolumnfalse}]

\section{Introduction}

One of the highlights of the past decade in the domain of high-energy scatterings 
concerns collective phenomena in small systems. It has been shown that high-multiplicity $pp$ events
show very similar collective features as earlier observed in heavy
ion collisions \cite{CMS:2010ifv}, see also the review \cite{Grosse-Oetringhaus:2024bwr}.
It is tempting to use the same theoretical tools (hydrodynamic evolution) as in "big systems",
and this is what I am going to do.

But here the applicability of viscous hydrodynamics has to be questioned. 
A possible way out is given in terms of
hydrodynamics attractors, whose existence has been shown in numerous publications 
\cite{Heller:2011,Heller:2015,Keegan:2015,Keegan:2016,Heller:2016,Romatschke:2017,Strickland:2017,Kurkela:2018,Strickland:2018,Giacalone:2019}.
In all cases, the time evolution is given in terms of a single scaling variable $\tilde{\omega}$, with very similar attractor curves,
showing a universal behavior at small and large $\tilde{\omega}$, where the latter refers to the late stage viscous hydro stage
 \cite{Giacalone:2019}. This is encouraging, but it should also be mentioned that these attractor studies employ strong assumptions 
(boost invariance, conformal symmetry, relaxation time approximation), so the applicability of viscous hydrodynamics remains an open question.

The present work is complementary to the above theoretical studies. It is often said that viscous hydro "works", but is this really true? 
I investigate in detail to what extent viscous hydro in the EPOS4 framework is compatible with experimental data, and I try to consider all kinds of observables. Many results have already been published \cite{werner:2023-epos4-overview,werner:2023-epos4-heavy,werner:2023-epos4-smatrix,werner:2023-epos4-micro,Zhao:2023,Zhao:2024}. Here I focus on "flow". 
It should be mentioned that Ref. \cite{Zhao:2023} already studied the question of "flow of charm",
but it has not been shown to what extent the "ordinary flow results" 
(more precisely multi-particle cumulants and flow harmonics) based on light-flavor hadrons
can be reproduced in this framework. This paper will fill the gap. 
 
In the EPOS4 approach, one distinguishes ``primary interactions''
and ``secondary interactions''. The former refer to parallel partonic
scatterings, happening at very high energies instantaneously at $t=0$,
such that any notion of a sequential ordering makes no sense. The
theoretical tool is S-matrix theory, using a particular form of the
S-matrix. The main new development in EPOS4 \cite{werner:2023-epos4-overview,werner:2023-epos4-heavy,werner:2023-epos4-smatrix,werner:2023-epos4-micro}
is a way to accommodate simultaneously: (1) rigorous parallel scattering,
(2) energy-momentum sharing, and (3) validity of the Abramovskii-Gribov-Kancheli
(AGK) theorem \cite{Abramovsky:1973fm}, which assures binary scaling
{[}in nucleus-nucleus ($AA$) scattering{]} and factorization \cite{Collins:1989}
{[}in proton-proton ($pp$) scattering{]} for hard processes, by introducing
(in a very particular way) saturation, compatible with recent ``low-x-physics''
considerations \cite{Gribov:1983ivg,McLerran:1993ka,kov95,kov96,jal97,jal97a,jal99a}. 

\begin{figure}
\begin{centering}
\includegraphics[scale=0.2]
{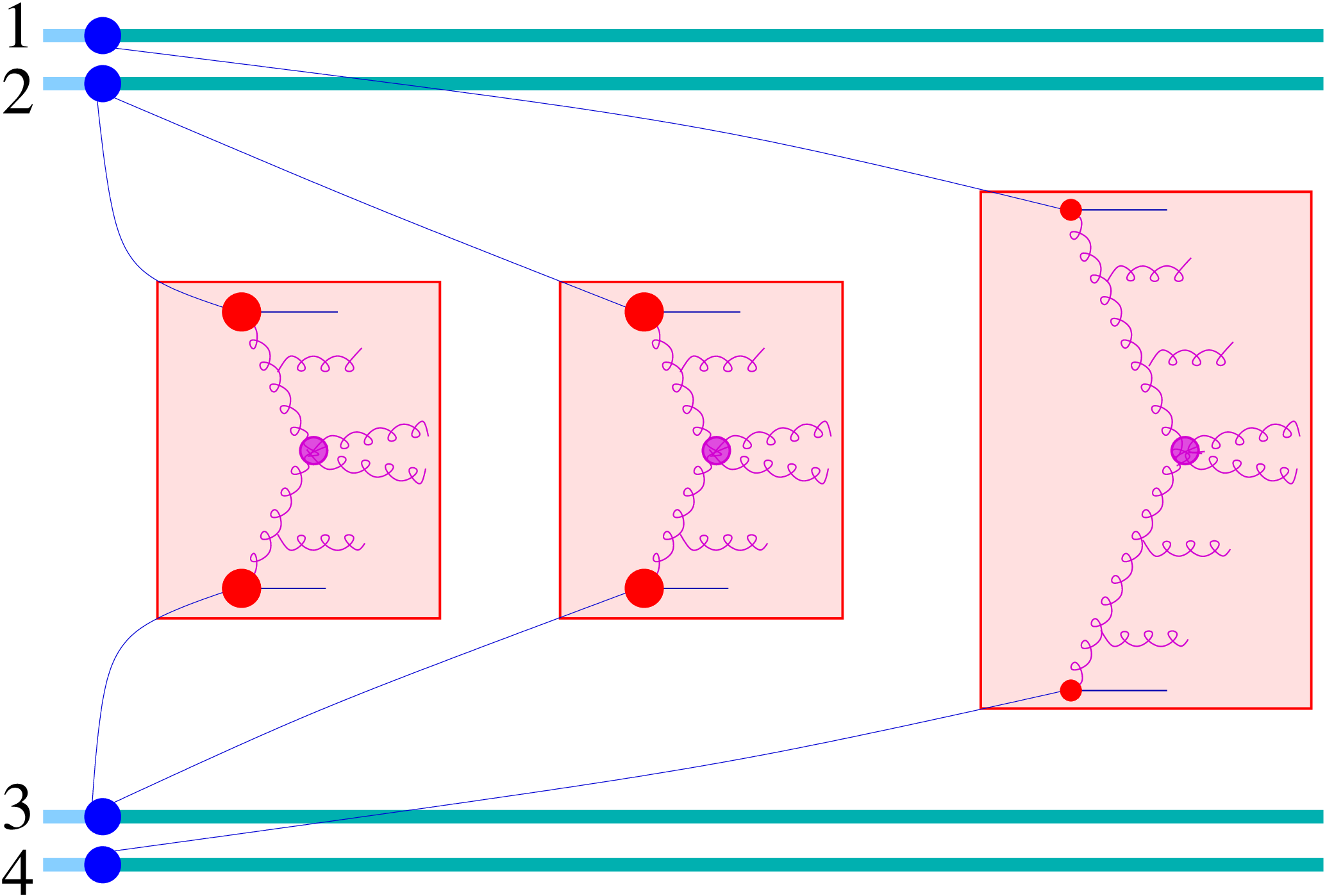}
\par\end{centering}
\caption{Sketch of the ``compensation'' of smaller energies (red box sizes)
by larger saturation scale values (red dots)\protect\label{Compensation-of-the},
in a collision of two nuclei with two nucleons each. }
\end{figure}

Although energy-momentum sharing makes things complicated, it is not
only mandatory for a consistent picture, it also allows one to understand
a crucial connection between factorization and saturation. The EPOS4
formalism \cite{werner:2023-epos4-overview,werner:2023-epos4-heavy,werner:2023-epos4-smatrix,werner:2023-epos4-micro}
is quite involved, but it is easy to understand the essential points,
see also the ``pedagogical presentation'' \cite{Werner:2024-pedagog}.
Let me qualitatively explain it by considering the scattering of two
nuclei ($A+B$ with $A=B=2$) with altogether 3 subscatterings, as
sketched in Fig. \ref{Compensation-of-the}. The two left scatterings
represent a double scattering involving the pair of nucleons 2 and
3, whereas the right one is just a single scattering between nucleons
1 and 4. Double scattering (per nucleon-nucleon pair) means energy
sharing, so the energy associated with each scattering is small (indicated
by small red boxes) compared to a single scattering. However, the
saturation scales associated with the double scatterings left are bigger
(indicated by big red dots) compared to the single scattering between
1 and 4, and correspondingly, the parton evolution is shorter due to the
bigger saturation scale. But \textbf{the central part responsible
for the hard scattering is identical} in all cases. This last point
is the crucial element, which assures that at the end the hard particle
production is independent of the number of scatterings, and therefore
the sum of all multiple scattering contributions is (up to a factor)
identical to the single scattering case. And this is what is needed
to get factorization (and binary scaling in nuclear scatterings) in
such a multiple scattering formalism.
All that means that one can do the same as models based on factorization
(defining and using parton distribution functions) to study hard processes
(this is needed to prove consistency), but one can do much more, and
this is very relevant when it comes to understanding high multiplicity
(= multiple scattering) $pp$ events. One of the highlights of the
past decade in our domain concerns actually collective phenomena in
small systems. It has been shown that high-multiplicity pp events
show very similar collective features as earlier observed in heavy
ion collisions \cite{CMS:2010ifv}. 

\begin{figure}[t]
\begin{centering}
\includegraphics[bb=0bp 0bp 567bp 520bp,clip,scale=0.38]
{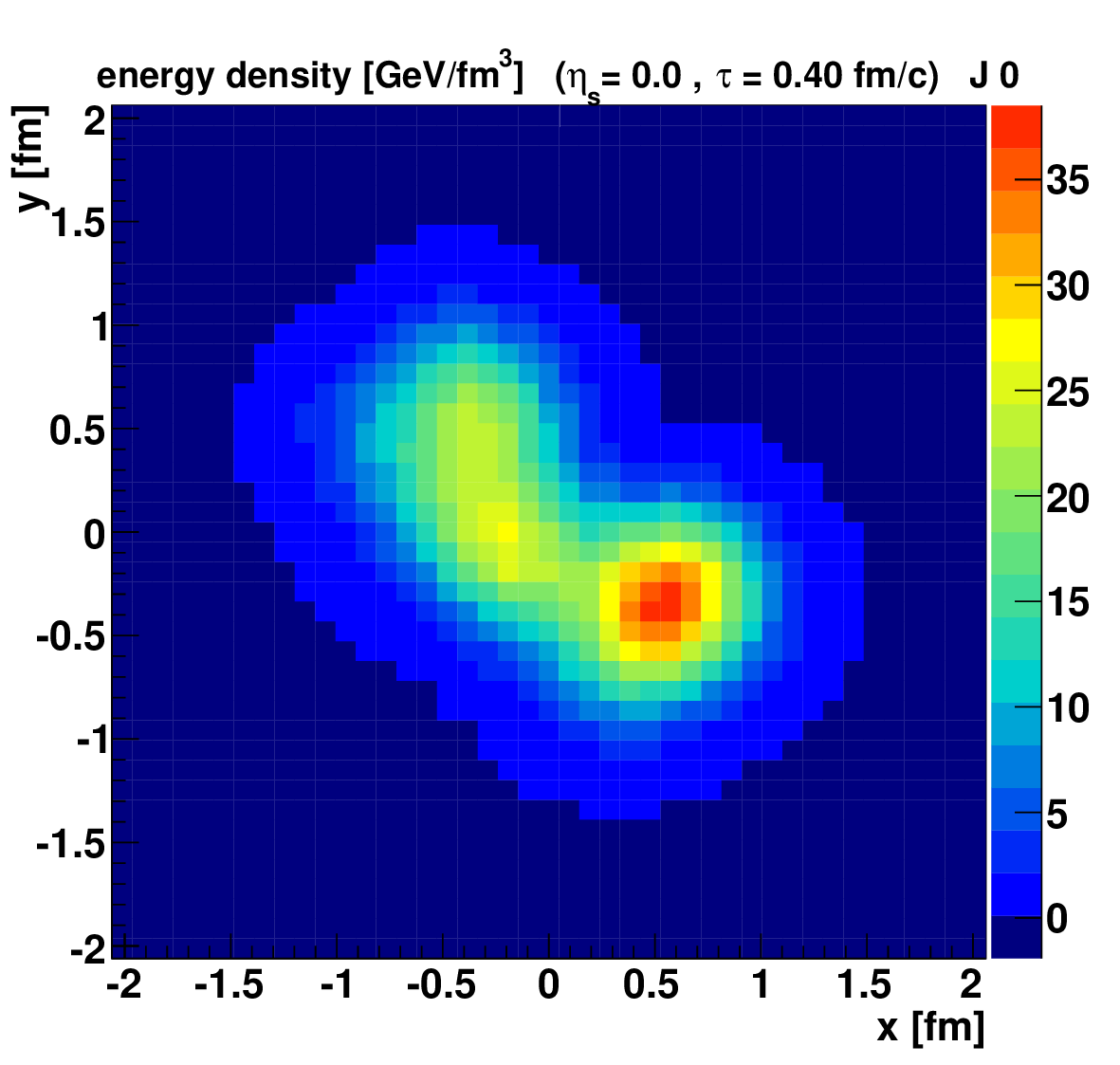}
\par\end{centering}
\begin{centering}
\includegraphics[bb=0bp 10bp 567bp 520bp,clip,scale=0.38]
{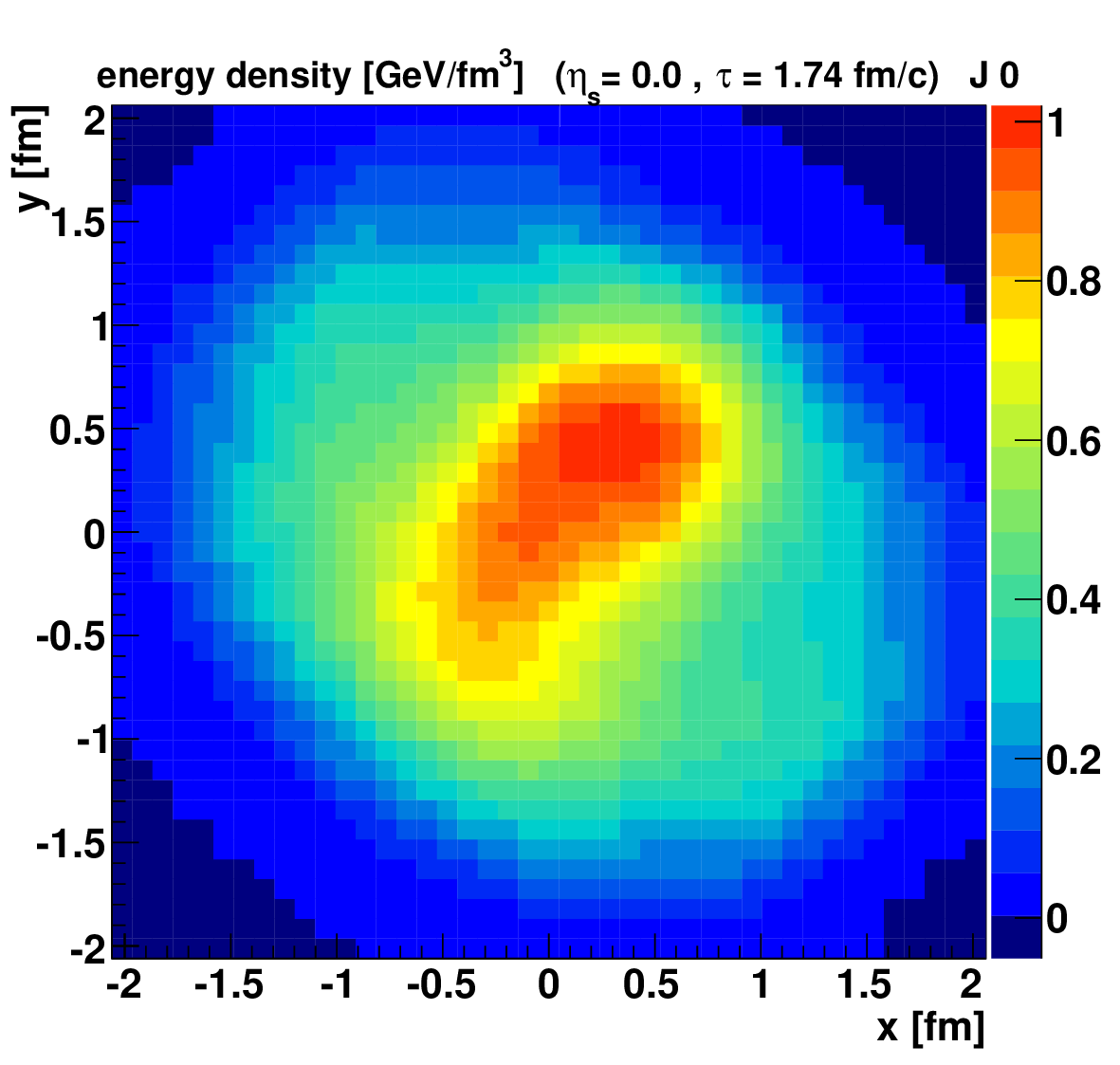}
\par\end{centering}
\caption{Energy density of the fluid (core) in the transverse plane ($x,y)$
for proton-proton scattering at 7 TeV involving 6 Pomerons. The upper
plot represents the start time $\tau_{0}$ (of the hydro evolution),
and the lower plot a later time $\tau_{1}$, close to the final freeze-out.
\protect\label{Energy-density-6-Pomerons}}
\end{figure}

\begin{figure}[h]
\begin{centering}
\includegraphics[bb=0bp 0bp 567bp 520bp,clip,scale=0.38]
{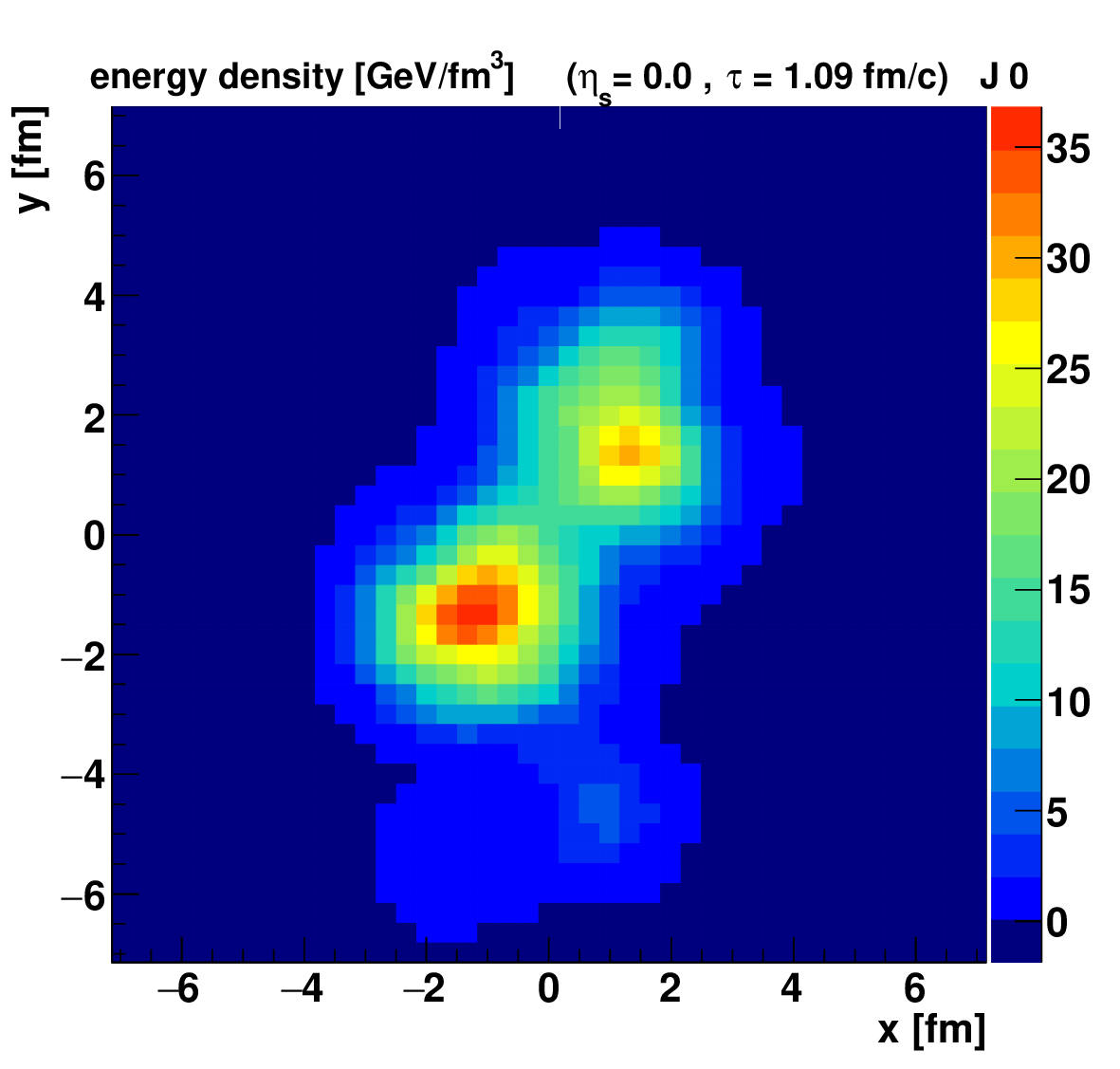}
\par\end{centering}
\centering{}\includegraphics[bb=0bp 10bp 567bp 520bp,clip,scale=0.38]
{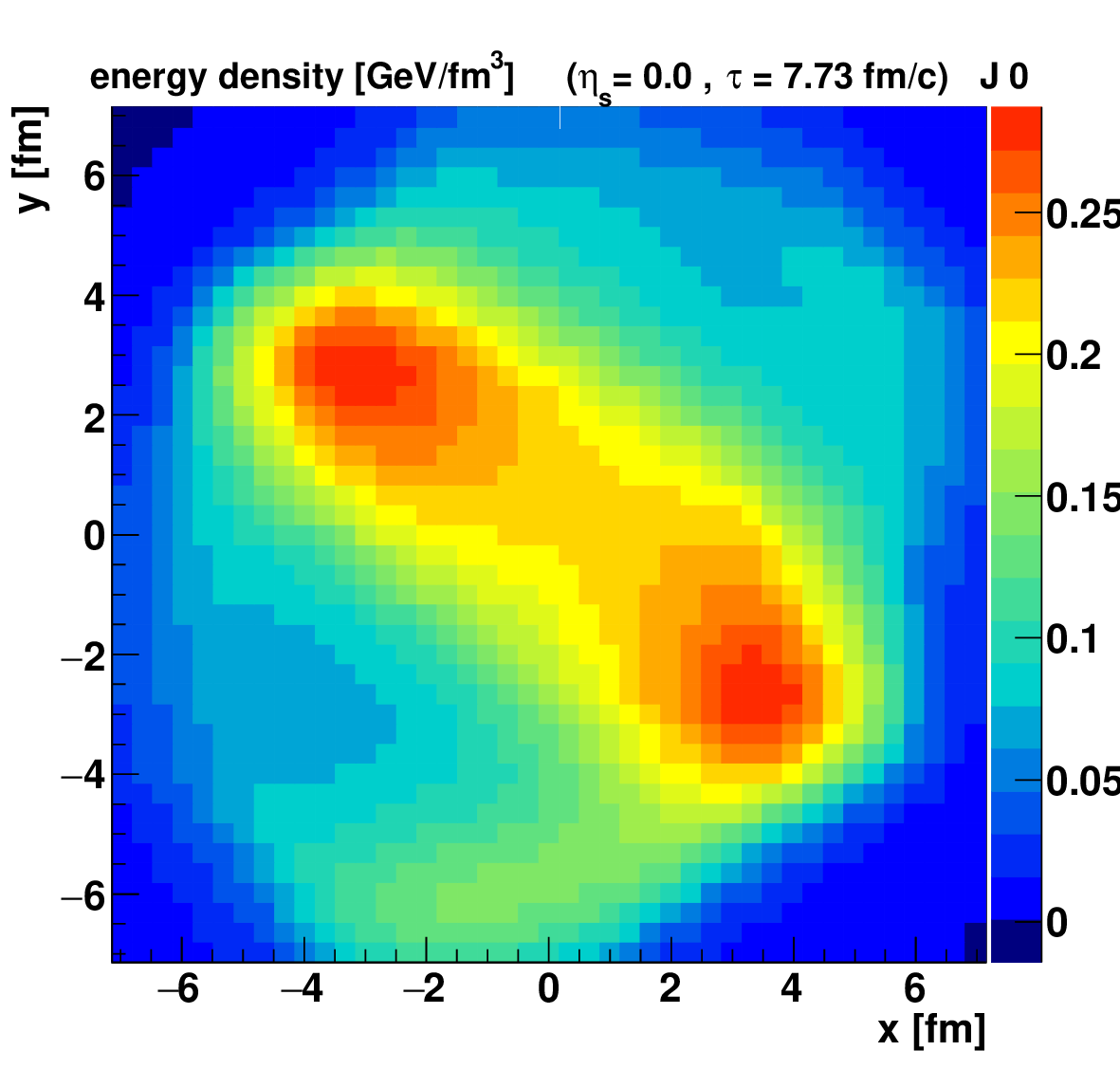}
\caption{Energy density of the fluid (core) in the transverse plane ($x,y)$
for a lead-lead scattering at 5.02 TeV with an impact parameter of
10.4 fm. The upper plot represents the start time $\tau_{0}$ (of
the hydro evolution), and the lower plot a later time $\tau_{1}$,
close to final freeze-out. \protect\label{Energy-density-lead-lead}}
\end{figure}

From the above-mentioned primary interactions, one obtains a more
or less important number of prehadrons. A core-corona procedure \cite{Werner:2007bf,Werner:2010aa,Werner:2013tya}
is employed (see Ref. \cite{werner:2023-epos4-micro} for the currently
employed procedure), where the prehadrons, considered at a given proper
time $\tau_{0}$, are separated into ``core'' and ``corona'' prehadrons,
depending on the energy loss of each prehadron when traversing the
``matter'' composed of all the others. Corona prehadrons (per definition)
can escape, whereas core prehadrons lose all their energy and constitute
what is called ``core'', which acts as an initial condition for
a hydrodynamic evolution \cite{Werner:2013tya,Karpenko_2014}. In
the following, in Figs. \ref{Energy-density-6-Pomerons} and \ref{Energy-density-lead-lead},
I will show the space-time evolution of typical events for small and
big systems.

Let me consider a (randomly chosen, but typical) 7 TeV proton-proton
scattering event with 6 Pomerons (representing roughly three times
the average). In Fig. \ref{Energy-density-6-Pomerons}, I plot the
energy density in the transverse plane ($x,y)$. I consider two snapshots,
namely at the start time of the hydro evolution $\tau_{0}=0.40\,\mathrm{fm/c}$
(upper plot) and a later time $\tau_{1}$ close to the final freeze-out
(lower plot). The initial distribution has an elongated shape (due
to the random positions of interacting partons). One can clearly see
that the final distribution is also elongated, but perpendicular
to the initial one, as expected in a hydrodynamical expansion. In
Fig. \ref{Energy-density-lead-lead}, I plot the energy density in
the transverse plane ($x,y)$ for a (randomly chosen, but typical)
5.02 TeV lead-lead scattering event with an impact parameter of 10.4
fm, again considering two snapshots, namely at the start time of the
hydro evolution $\tau_{0}=0.40\,\mathrm{fm/c}$ (upper plot) and a
later time $\tau_{1}$ close to final freeze-out (lower plot). The
lead-lead plots look very similar to what has been seen before for
proton-proton scattering. Also here (for lead-lead), one can see that
the final distribution is elongated, perpendicular to the initial
one. 

The Figs. \ref{Energy-density-6-Pomerons} and \ref{Energy-density-lead-lead}
are just two examples, but looking at more cases, one can see that
the space-time evolution of the energy density in proton-proton scattering
and lead-lead scattering look very similar. In case of an elongated
initial shape, one gets at the end as well an elongated shape, but
perpendicular to the initial one, as expected from a hydrodynamic
expansion, and which should eventually lead to nonvanishing values
for the elliptical flow $v_{2}$.

But before coming to the discussion of flow harmonics $v_{n}$, based
on the momenta of the final particles, one has to mention how to get
from a fluid (as shown in the Figs. \ref{Energy-density-6-Pomerons}
and \ref{Energy-density-lead-lead}) to hadrons. The evolution of
the core ends whenever the energy density falls below some critical
value $\epsilon_{\mathrm{FO}}$, which marks the point where the fluid
``decays'' into hadrons. It is not a switch from fluid to particles,
it is a sudden decay, called ``hadronization''. In EPOS4, as discussed
in detail in \cite{werner:2023-epos4-micro}, a new procedure was
developed for computing energy-momentum flow through the ``freeze-out (FO) hypersurface''
defined by $\epsilon_{\mathrm{FO}}$, which allows defining an effective
invariant mass, which decays according to microcanonical phase space
into hadrons, which are then Lorentz boosted according to the flow
velocities computed at the FO hypersurface. Also were developed new
and very efficient methods for the microcanonical procedure \cite{werner:2023-epos4-micro}.
Also in the full scheme, including primary and secondary interactions,
energy-momentum and flavors are conserved.

\section{Elliptical flow in \emph{pp} collisions at 13 TeV, using the symmetric
scenario}

In Ref. \cite{ATLAS:2017harmonics}, the ATLAS collaboration studies
multi-particle cumulants and corresponding Fourier harmonics (or flow
harmonics) for small collision systems ($pp$ collisions at $\sqrt{s_{NN}}$
= 5.02 TeV) and compares the results with those for Pb+Pb collisions
at $\sqrt{s_{NN}}$ = 2.76 TeV at low multiplicity. In all cases,
cumulants and flow harmonics are presented as a function of the charged
particle multiplicity $N_{\mathrm{ch}}$, defined in the same way
for the different systems (referred to as $\left\langle N_{\mathrm{ch}}(p_{\mathrm{T}}>0.4)\right\rangle $in
Ref. \cite{ATLAS:2017harmonics}. 

Popular procedures to determine flow harmonics are the cumulant method%
{} \cite{Bilandzic:2010-cumulants,ATLAS:2014-PbPb3-flow}, based on
\cite{Borghini:2000sa,Borghini:2001vi,Borghini:2001zr}, and the scalar
product method%
{} \cite{STAR:2002-Elliptic,ALICE:2014-Elliptical} (see also App. \ref{_______Computing-flow-harmonics_______}),
where the former is employed in Ref. \cite{ATLAS:2017harmonics}.
In addition, a requirement of a peudorapidity gap |\ensuremath{\Delta}\textgreek{\texteta}|
> 2 may be implemented in calculating the 2 particle cumulants, in
order to reduce the so-called nonflow effects.

In Fig 9 of Ref. \cite{ATLAS:2017harmonics}, one compares the multiplicity
dependence of the elliptical flow $v_{2}$ (with a pseudorapidity
gap of 2 units) for different systems, showing a completely different
behavior. Whereas for PbPb one observes a strong increase, the $pp$
results are constant. The heavy ion curve is (at least qualitatively)
easy to understand. An important ``source'' for creating elliptical
flow is the initial elliptical shape of the overlap zone, in semiperipheral
collisions. Going to very peripheral ion-ion collisions, the overlap
zone vanishes, only very peripheral nucleons participate in the scattering,
which leads to a decrease of the elliptical flow. A constant elliptical
flow in $pp$ is not so easy to understand, so I will investigate
this in the following. 

\begin{figure}[t]
\centering{}\includegraphics[bb=30bp 50bp 820bp 530bp,clip,scale=0.32]
{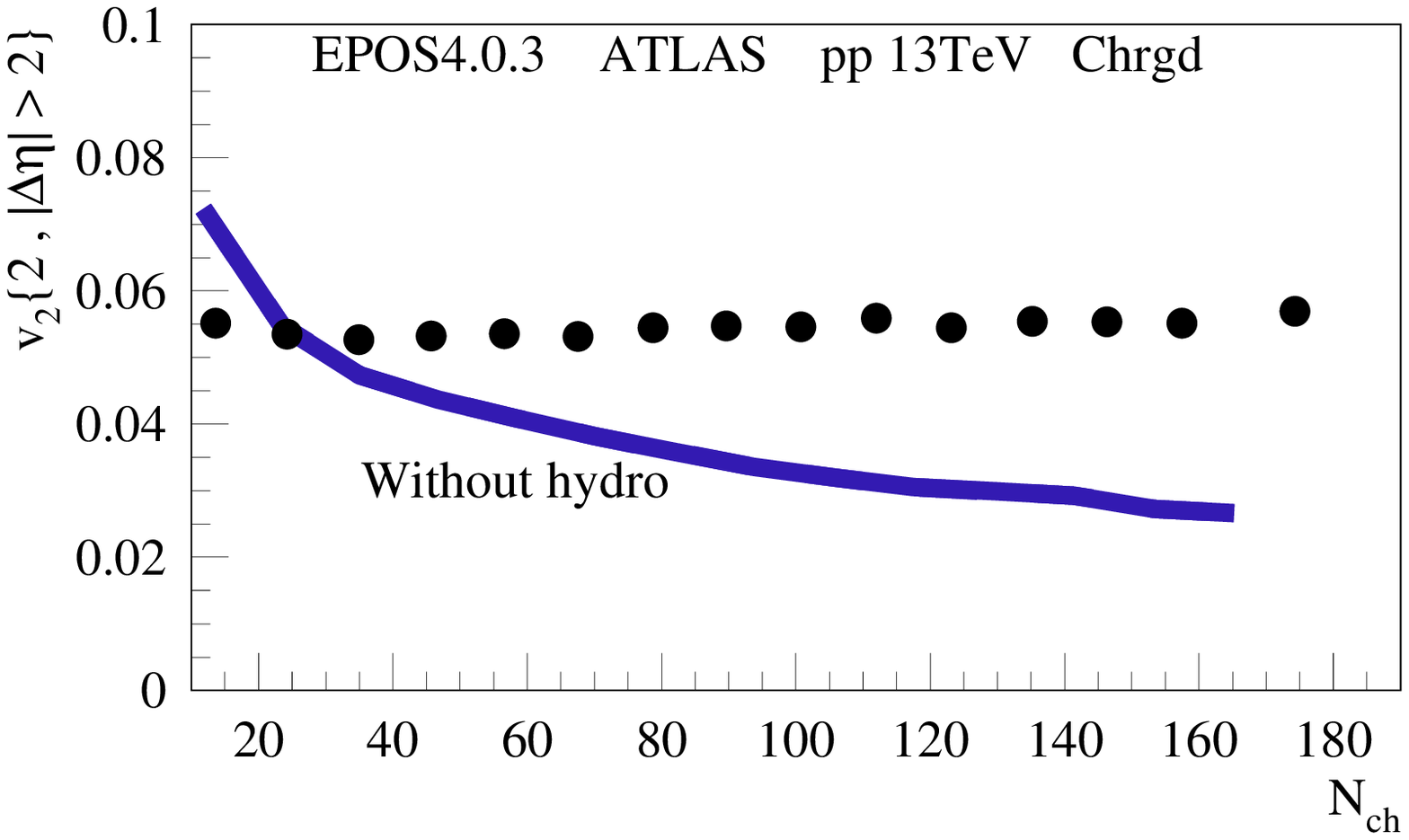}
\caption{Elliptical flow $v_{2}\{2,|\Delta\eta|>2\}$, shown as a function
of the multiplicity $N_{\mathrm{ch}}$, in pp collisions at 13 TeV.
Compared are  the simulations without hydro (blue line) with data
from ATLAS \cite{ATLAS:2017harmonics} (black points). \protect\label{sel10}}
\end{figure}

Let me first discuss results from EPOS4 simulations without hydrodynamical
evolution, i.e., only primary scatterings. In Fig. \ref{sel10}, I
plot the corresponding elliptical flow $v_{2}\{2\}$ with a pseudorapidity
gap of 2 units, in short $v_{2}\{2\,,|\Delta\eta|>2\}$, as a function
of the multiplicity $N_{\mathrm{ch}}$, in $pp$ collisions at 13
TeV (blue line), compared with the data from ATLAS \cite{ATLAS:2017harmonics}
(black points). The theoretical results represent a so-called "nonflow"
contribution, because per construction there is no fluid involved,
so the correlations have a different origin, like dijet production. As
expected, the curve drops with increasing multiplicity. One can see
that despite the application of a rapidity gap of two units, meant
to reduce nonflow effects, the latter are significant at low multiplicity.

Let me now consider the full simulation, including hydrodynamic evolution
and hadronic cascade. An important issue for creating azimuthal asymmetries
is a corresponding asymmetry in the initial matter distribution,
where the latter is determined by the positions in the transverse
plane of the colliding partons. Traditionally (as realized in EPOS3),
the transverse positions $\vec{b}_{i}$ of the partons $i$ relative
to the nucleon center are generated as 
\begin{equation}
\vec{b}_{i}=b_{i}(\cos\varphi_{i},\sin\varphi_{i}),\label{symmetric=000020scenario}
\end{equation}
where the $\varphi_{i}$ are randon angles in $[0,2\pi]$ and the
$b_{i}$ are random numbers generated according to some simple law
$f(b)$, for $1\le i\le N$ in case of multiple ($N$-fold) scattering.
Since the laws are perfectly symmetric, one refers to this approach
as ``symmetric scenario''. The only source of asymmetery is the
randomness in the event-by-event treatment, but there is no geometric
origin as in the case of heavy ion collisons. For completeness it should be said that there is some geometrical effect due to finite impact parameters, but for the (interesting) multiple scattering events, the impact parameters are very small.

Before discussing simulation results, one needs to overcome some technical
challenges. In Fig. \ref{sel10}, results are presented up to multiplicies
of $N_{\mathrm{ch}}=190$. But doing now full simulations, including
hydrodynamic evolutions, it becomes impossible to go so far. As shown
in App. \ref{_______High=000020multiplicity=000020events_______-1},
one may use the number of Pomerons as a trigger to generate high multiplicity
events. I will therefore do minimum bias simulations, but show results
only up to $N_{\mathrm{ch}}=100$, and then simulate events with more
than 12 Pomerons, but show results only beyond $N_{\mathrm{ch}}=100$.
This method helps, because the first part of the simulation (to determine
the number of Pomerons) is fast. The following results will be shown
as ``broken lines'', with the different pieces corresponding to
different triggers.

\begin{figure}[t]
\centering{}\includegraphics[bb=17bp 50bp 820bp 530bp,clip,scale=0.32]
{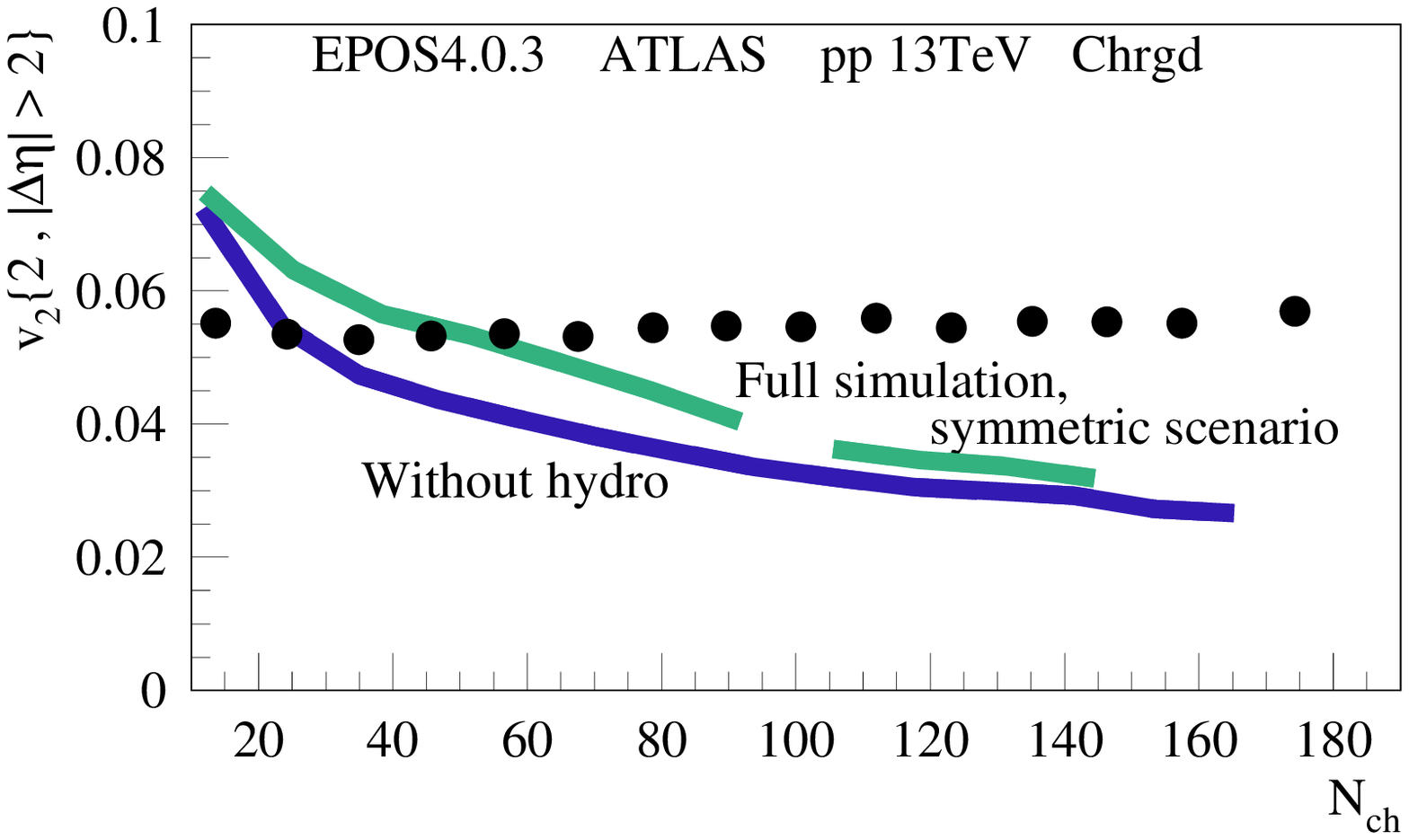}
\caption{Same as Fig. \ref{sel10}, but in addition the full simulation, using
the symmetric scenario (green broken line). \protect\label{sel11}}
\end{figure}

In Fig. \ref{sel11}, I consider the full simulation, using the ``symmetric
scenario'', and plot the corresponding elliptical flow $v_{2}\{2,|\Delta\eta|>2\}$,
as a function of the multiplicity (green broken line). The elliptical
flow for the full simulation (including hydro) is bigger than the
one without hydro, but the curve is not at all flat like the ATLAS data,
but it quickly decreases with multiplicity, well below the data points.
It even approaches the (blue) nonflow curve at large multiplicity.

Such a failure is actually expected in the symmetric scenario. As
already said, the laws of probability concerning the positions in
the transverse plane of the colliding partons are symmetric, and an
asymmetry only occurs due to the randomness of the event-by-event
generation. But increasing multiplicity is correlated with an increasing
number of scatterings, having an increasing number of partons (and
eventually strings) involved, which makes the initial matter distribution
more and more symmetric. And a hydrodynamical evolution based on symmetric
initial matter distributions will not produce any $v_{2}$. It is therefore
unavoidable, in the multiple scattering formalism, using this "symmetric
scenario", to have $v_{2}$ vanishing at large multiplicity. In other
words, it is impossible to get a flat curve as seen in the ATLAS data.

\section{Elliptical flow in \emph{pp} collisions at 13 TeV, using the dipole
scenario}

Already in EPOS4.0.0, the possibility was introduced to change the
laws for generating the tranverse positions of the partons. Instead
of Eq. (\ref{symmetric=000020scenario}), one may use
\begin{equation}
\vec{b}_{i}=b_{i}(\cos\varphi_{i},\sin\varphi_{i})\pm\frac{b_{\mathrm{dipole}}}{2}(\cos\varphi_{\mathrm{dipole}},\sin\varphi_{\mathrm{dipole}}),\label{dipole=000020scenario}
\end{equation}
where the ``dipole angle'' $\varphi_{\mathrm{dipole}}$ is chosen
randomly in $[0,2\pi]$, but remains the same in case of multiple
scatterings. In other words, the partons are now generated around
two centers. I refer to this picture as ``dipole scenario''. The
dipole orientations are random, but it happens that in a proton-proton
scattering, i.e., the scattering of two dipoles, the orientations
are (anti)parallel, creating an elongated initial matter distribution.
Therefore, in addition to randomness as an origin of asymmetry, one
has here as well a geometric element. This dipole picture was never
systematically studied in EPOS4.0.0 publications, the present paper
should fill the gap.

\begin{figure}[b]
\centering{}\includegraphics[bb=17bp 50bp 820bp 530bp,clip,scale=0.32]
{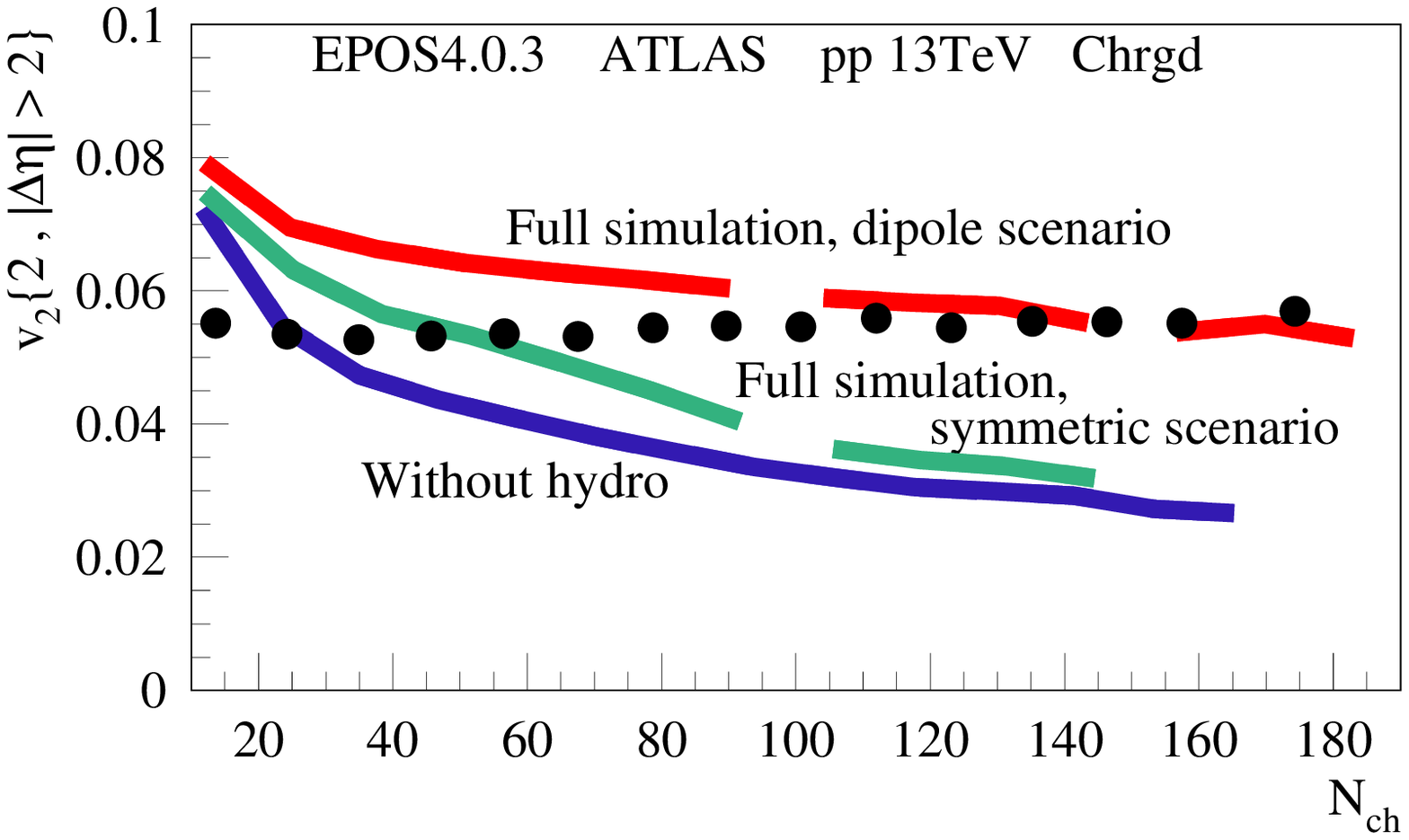}
\caption{Same as Fig. \ref{sel11}, but in addition the full simulations, using
the dipole scenario (red broken line). \protect\label{sel12}}
\end{figure}

In Fig. \ref{sel12}, I consider the full simulation, using the ``dipole
scenario'', and plot the corresponding elliptical flow $v_{2}\{2,|\Delta\eta|>2\}$,
as a function of the multiplicity (red broken line). For small $N_{\mathrm{ch}}$,
the red curve exceeds the data, it seems that the nonflow is somewhat
too big. But beyond a multiplicity of 100, one observes an essentially
flat curve, as in the experimental data. Such a behavior is expected,
since an increasing number of partons (and strings) does not change
the dipole form of the matter distribution. Whereas the flatness of
the curve for large multiplicities is a real feature of the approach,
its value depends on the parameter $b_{\mathrm{dipole}}$. I take
a dipole size $b_{\mathrm{dipole}}$ of 1.5 fm.

It is essential that one has here a geometrical feature (the dipole)
compared to randomeness as the origin of a possible asymmetry in the
symmetric scenario, because the latter leads unavoidably to a continuous
decrease of the elliptical flow with increasing multiplicity. 

\begin{figure}[t]
\centering{}\hspace*{-0.2cm}\includegraphics[bb=17bp 30bp 595bp 680bp,clip,scale=0.5]
{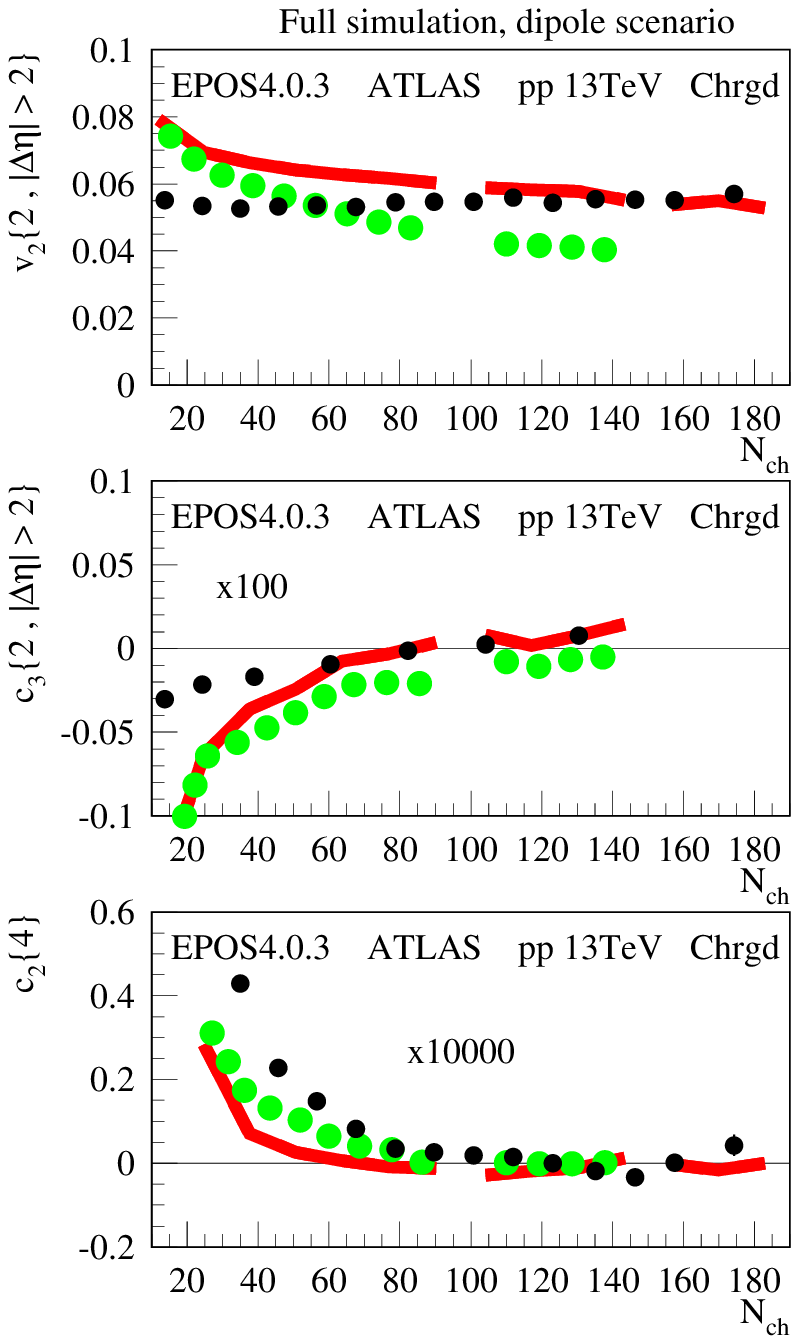}
\caption{Flow harmonics and cumulants versus the multiplicity $N_{\mathrm{ch}}$,
in pp collisions at 13 TeV. One compares the full simulations, using
the dipole scenario (red lines), with data from ATLAS \cite{ATLAS:2017harmonics}
(black points). Also shown: results for shear viscosity over entropy density ($\eta/s$) of 0.24 (light green dotted). \protect\label{sel7}}
\end{figure}

In the above-mentioned ATLAS paper, they present various results concerning
the multiplicity dependence of cumulants and flow harmonics. In Fig.
\ref{sel7}, I show in addition to the already discussed $v_{2}\{2,|\Delta\eta|>2\}$
(upper panel), the cumulants $c_{3}\{2,|\Delta\eta|>2\}$ (middle
panel), and $c_{2}\{4\}$ (lower panel. For the definitions, see App.
\ref{_______Computing-flow-harmonics_______}. The data are presented
as black dots. Also shown (as red lines) are the EPOS4 results for
the full simulation using the dipole scenario. In all cases, for multiplicities
beyond 100, the simulations follow the trend of the data. 
Concerning the hydro evolution, the default value of shear viscosity over entropy density ($\eta/s$) of 0.08 is used for all plots. 
In Fig. \ref{sel7}, in addition results  for $\eta/s=0.24$ are shown (light green dotted). 
At large multiplicity, they are parallel to the curves for $\eta/s=0.08$, but lower, which would require a larger dipole size to be close to the data. 

\begin{figure}[t]
\centering{}\hspace*{-0.2cm}\includegraphics[bb=17bp 30bp 595bp 680bp,clip,scale=0.5]
{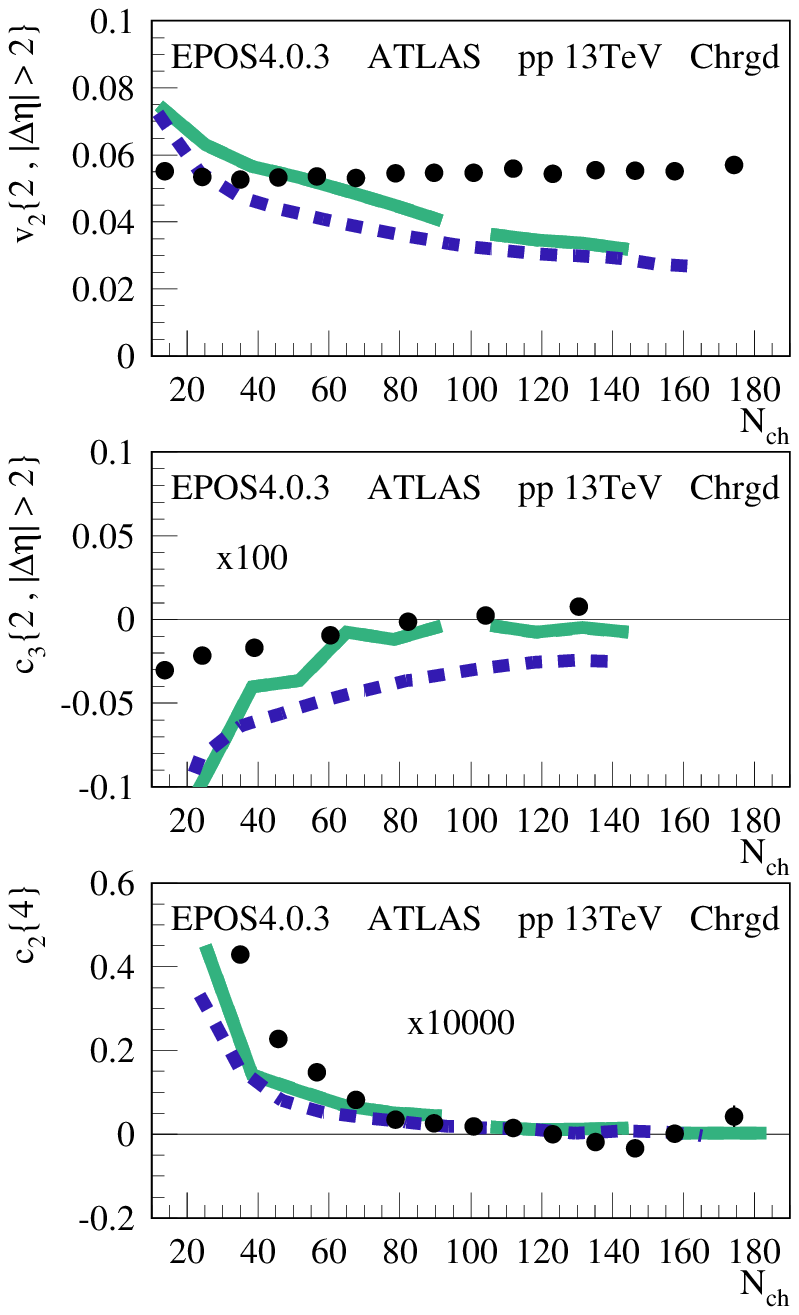}
\caption{Cumulants and flow harmonics versus the multiplicity $N_{\mathrm{ch}}$,
in pp collisions at 13 TeV. One compares the full simulations, using
the symmetric scenario (green lines) and the simulations without hydro
(blue dashed lines) with data from ATLAS \cite{ATLAS:2017harmonics}
(black points). \protect\label{sel98}}
\end{figure}

In order to see the effect due to the dipole feature, I plot
in Fig. \ref{sel98}  the corresponding quantities based on full simulations
using the symmetric scenario (green lines), and also the ones for
simulations without hydro, i.e., the nonflow case (blue dashed lines). 
The upper panel showns  $v_{2}\{2,|\Delta\eta|>2\}$, 
with both curves dropping below the data at large multiplicity, as already seen earlier, 
and here the dipol geometry helps to get close to the data, as shown in Fig. \ref{sel7}.

Concerning the cumulants $c_{3}\{2,|\Delta\eta|>2\}$ (middle panel),
there is a big negative nonflow effect (blue dashed line), getting
smaller (but still significant) at larger multiplicities (one cannot
go beyond 140 due to limited statistics). The full simulation employing
the symmetric scenario (green line) seems to compensate for the negative
nonflow, to give a vanishing cumulant. Comparing this result with
$c_{3}\{2,|\Delta\eta|>2\}$ for the dipole scenario in Fig. \ref{sel7},
one can see that there is no "dipole effect", which is understandable:
The dipole geometry does not help to create triangular initial shapes,
the latter are purely random. An important point here: a vanishing
cumulant at intermediate multiplicity is not trivial, it is the result
of two effects (nonflow and flow) which compensate each other. 

Finally, the four-particle cumulants $c_{2}\{4\}$ (lower panel) for
the simulations without hydro (blue dashed line) and those of the
symmetric scenario (green line) are roughly zero beyond multiplicities
of 80, close to the data. The same can be said concerning the dipole
scenario shown in Fig. \ref{sel7}.

\section{Heavy-ion results}

\begin{figure}
\centering{}\hspace*{-0.2cm}\includegraphics[bb=17bp 30bp 595bp 680bp,clip,scale=0.51]
{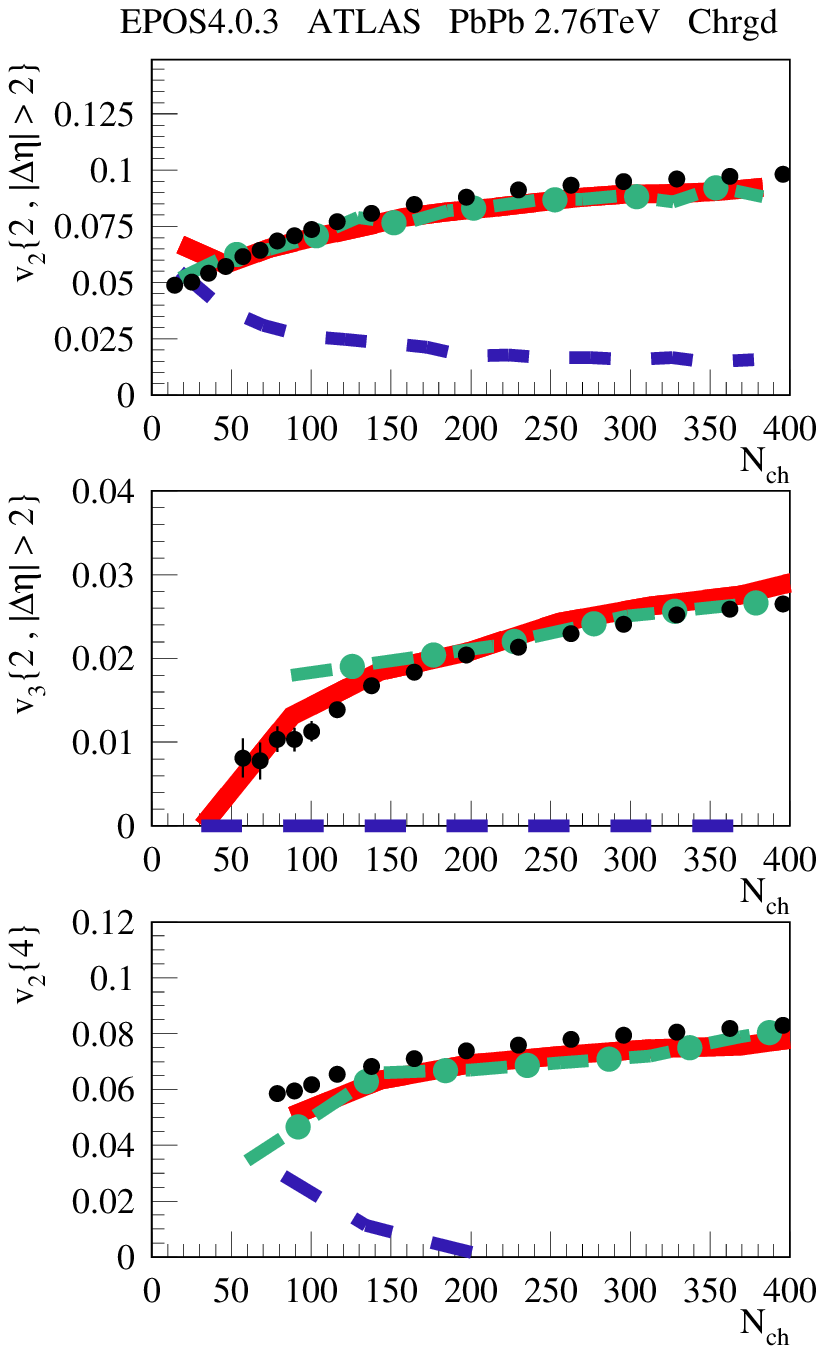}
\caption{Different flow harmonics versus the multiplicity $N_{\mathrm{ch}}$,
in PbPb collisions at 2.76 GeV. One compares the full simulations
using the dipole scenario (red lines), the full simulations using
the symmetric scenario (green dashed-dotted lines), and the simulations
without hydro (blue dashed lines), with data from ATLAS \cite{ATLAS:2017harmonics}
(black points)\label{sel4}. }
\end{figure}

\begin{figure}
\centering{}\hspace*{-0.2cm}\includegraphics[bb=120bp 30bp 595bp 520bp,clip,scale=0.48]
{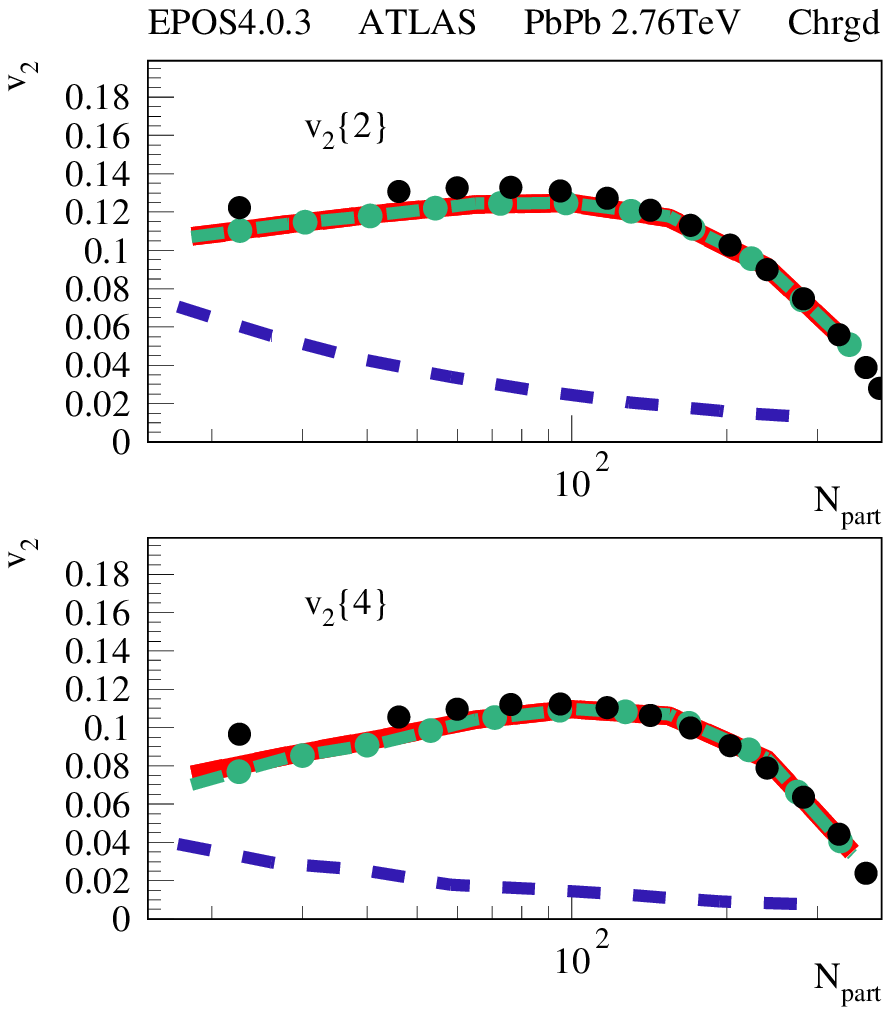}
\caption{Flow harmonics $v_{2}\{2\}$ and $v_{2}\{4\}$ versus $N_{\mathrm{part}}$,
in PbPb collisions at 2.76 GeV. One compares the full EPOS4 simulations
using the dipole scenario (red lines), the symmetric scenario (green
dashed dotted lines), and the simulations without hydro (blue dashed
lines), with data from ATLAS \cite{ATLAS:2014-PbPb3-flow} (black
points)\label{sel3-1}.}
\end{figure}

\begin{figure}
\centering{}\hspace*{-0.2cm}\includegraphics[bb=120bp 30bp 595bp 520bp,clip,scale=0.48]
{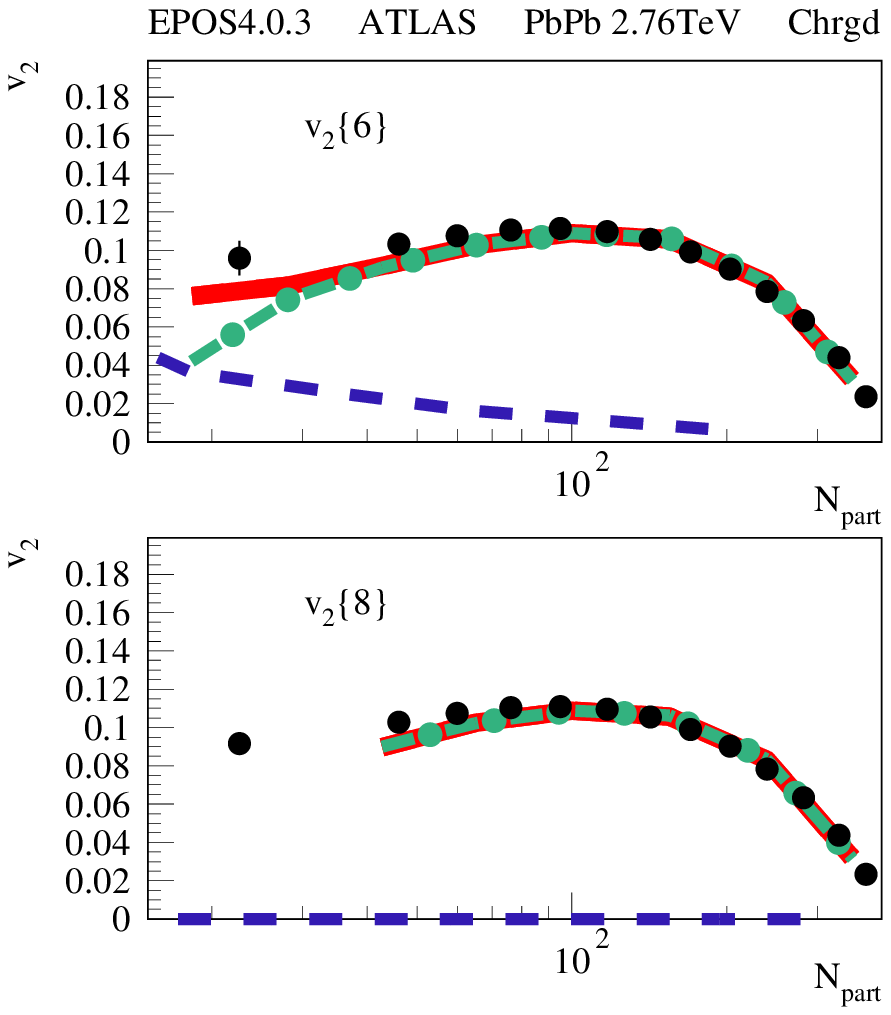}
\caption{Flow harmonics $v_{2}\{6\}$ and $v_{2}\{8\}$ versus $N_{\mathrm{part}}$,
in PbPb collisions at 2.76 GeV. One compares the full EPOS4 simulations
using the dipole scenario (red lines), the symmetric scenario (green
dashed dotted lines), and the simulations without hydro (blue dashed
lines), with data from ATLAS \cite{ATLAS:2014-PbPb3-flow} (black
points)\label{sel3-2}.}
\end{figure}

\begin{figure}
\centering{}\hspace*{-0.2cm}\includegraphics[bb=120bp 30bp 595bp 520bp,clip,scale=0.48]
{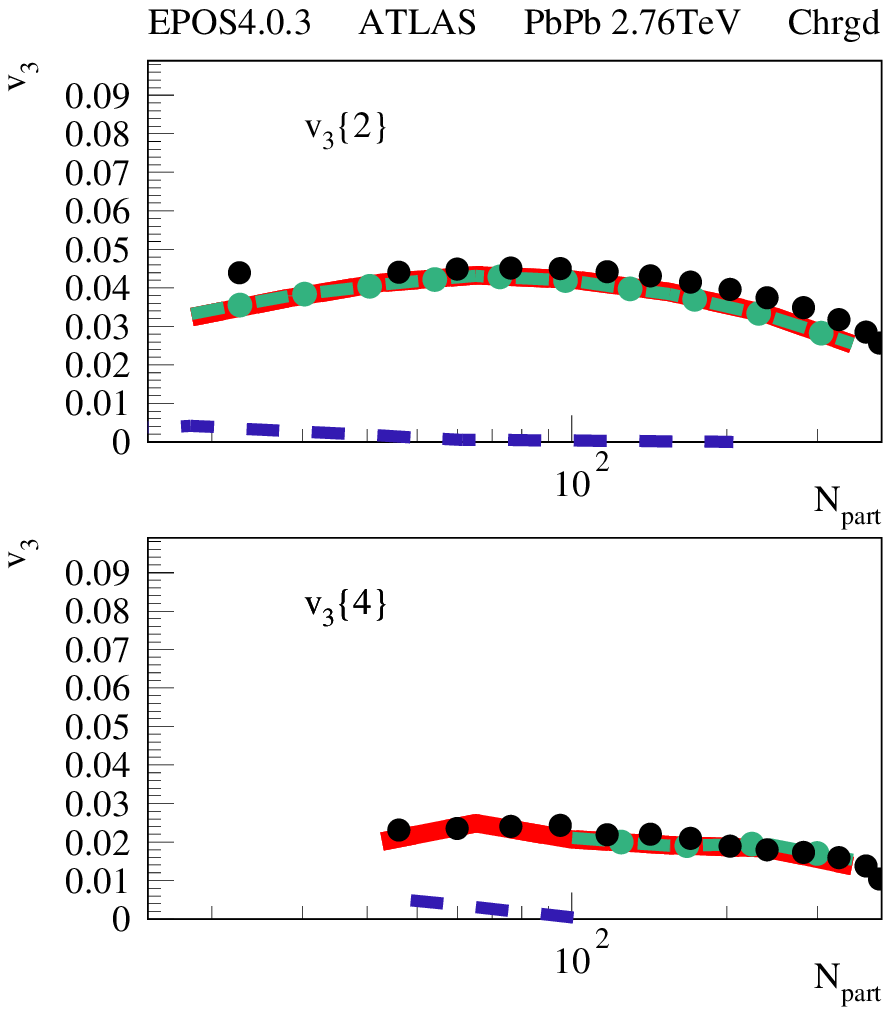}
\caption{Flow harmonics $v_{3}\{2\}$ and $v_{3}\{4\}$ versus $N_{\mathrm{part}}$,
in PbPb collisions at 2.76 GeV. One compares the full EPOS4 simulations
using the dipole scenario (red lines), the symmetric scenario (green
dashed dotted lines), and the simulations without hydro (blue dashed
lines), with data from ATLAS \cite{ATLAS:2014-PbPb3-flow} (black
points)\label{sel3-3}.}
\end{figure}

\begin{figure}
\centering{}\hspace*{-0.2cm}\includegraphics[bb=120bp 30bp 595bp 520bp,clip,scale=0.48]
{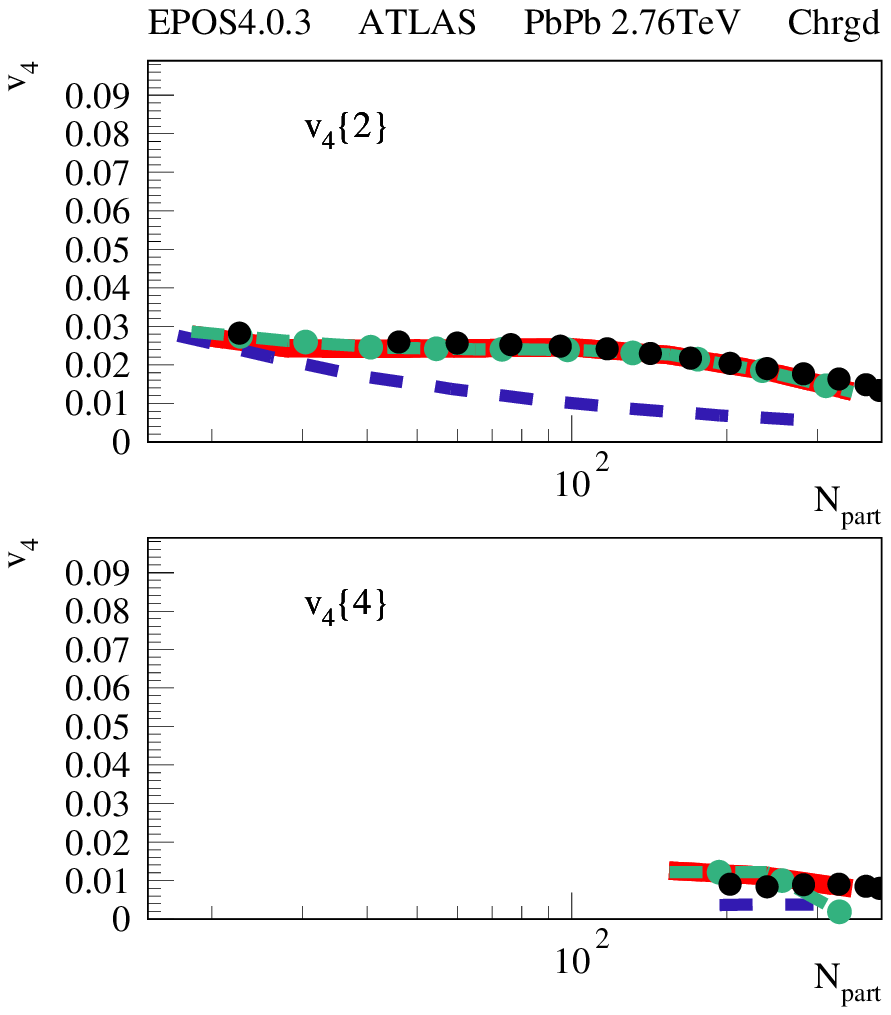}
\caption{Flow harmonics $v_{4}\{2\}$ and $v_{4}\{4\}$ versus $N_{\mathrm{part}}$,
in PbPb collisions at 2.76 GeV. One compares the full EPOS4 simulations
using the dipole scenario (red lines), the symmetric scenario (green
dashed dotted lines), and the simulations without hydro (blue dashed
lines), with data from ATLAS \cite{ATLAS:2014-PbPb3-flow} (black
points) \label{sel3-4}.}
\end{figure}

In Figs. \ref{sel4}, \ref{sel3-1}, \ref{sel3-2}, \ref{sel3-3}, and \ref{sel3-4}, I will compare experimental
data on flow harmonics in PbPb collisions at 2.76 TeV with full EPOS4
simulations, in the dipole scenario, which is the default in EPOS4.0.3,
with identical dipole sizes in the case of proton\textendash proton and
Pb+Pb collisions. I will also show results for the symmetric scenario
and for simulations without hydro.

In Ref. \cite{ATLAS:2017harmonics}, not only $pp$ scattering was
studied (discussed earlier), but as well low multiplicity Pb+Pb collisions,
with the same definition for the multiplicity $N_{\mathrm{ch}}$ in
both cases. In Fig. \ref{sel4}, I show $v_{2}\{2,|\Delta\eta|>2\}$
(upper panel), $v_{3}\{2,|\Delta\eta|>2\}$ (middle panel), and $v_{2}\{4\}$
(lower panel). For the definitions, see App. \ref{_______Computing-flow-harmonics_______}. 
The upper panel shows the elliptical flow (with $\eta$ gap), i.e.,
the same quantity as shown in the upper panel of Fig. \ref{sel98}
for pp scattering. Whereas the latter shows a flat behavior, the heavy
ion result shows up as a monotonically increasing curve, with almost
no difference between the dipole and the symmetric scenario, both
being close to the data.
The fundamental difference between proton-proton and heavy-ion collisions is
the fact that increasing multiplicity in the case of proton-proton is
due to multiple scatterings involving the same projectile and target
nucleon (there is only one), whereas in heavy-ion scattering, the
number of participating nucleons increases. Whereas in proton-proton
there is a privileged situation where both dipoles are parallel,
which counts for all multiple scatterings, in heavy-ion collisions
there is a large number of nucleon-nucleon pairs involved, and it
is very unlikely that all of them are parallel to each other.
As shown by the blue dashed curve, there is also a nonflow effect
at small multiplicities, as in proton-proton.


The middle panel of Fig. \ref{sel4} shows triangular flow $v_{3}\{2,|\Delta\eta|>2\}$,
with a similar behavior compared to the $v_{2}$ result: One can see
a monotonically increasing curve, with almost no difference between
the dipole and the symmetric scenario, both being close to the data.
The absolute size of the triangular flow is smaller than the one for
the elliptical flow, because there is no geometric origin for the
former.

The elliptical flow (based on four particle correlations) $v_{2}\{4\}$
(lower panel) shows a similar behavior to the two-particle result
$v_{2}\{2,|\Delta\eta|>2\}$, and again there is almost no difference
between the dipole and the symmetric scenario, both being close to
the data.

In Figs. \ref{sel3-1} - \ref{sel3-4}, the focus is on flow harmonics
based on multiple particle cumulants, as a function of the centrality
in terms of the number of participants $N_{\mathrm{part}}$, in PbPb
collisions at 2.76 GeV. I compare ATLAS data \cite{ATLAS:2014-PbPb3-flow}
with the full EPOS4 simulations using the dipole scenario (red lines),
the full simulations using the symmetric scenario (green dashed dotted
lines), and the simulations without hydro (blue dashed lines). The
range of $N_{\mathrm{part}}$ from 20 to 400 covers all centralities,
peripheral ones on the left and central ones on the right. 

In Fig. \ref{sel3-1}, elliptical flow results $v_{2}\{2\}$ and $v_{2}\{4\}$
are shown, based on two and four-particle cumulants. The nonflow effect
(blue dashed curves) is smaller in the four-particle case. Again,
there is almost no difference between the dipole and the symmetric
scenario; both are close to the data, and in both cases the elliptical
flow increases up to roughly $N_{\mathrm{part}}=100$, then decreases.
The increase in the case of four-particle cumulants is more pronounced. 

In Fig. \ref{sel3-2}, elliptical flow results $v_{2}\{6\}$ and $v_{2}\{8\}$
are shown, based on six and eight particle cumulants, with results
being similar to the two and four-particle ones. 

Also for $v_{3}\{2\}$, $v_{3}\{4\}$, $v_{4}\{2\}$, and $v_{4}\{4\}$,
as shown in Fig. \ref{sel3-3} and \ref{sel3-4}, there is almost
no difference between the dipole and the symmetric scenario, and both
are close to the data.

\section{More results \textendash{} for completeness}

\begin{figure}
\hspace*{-0.4cm}\includegraphics[bb=30bp 40bp 595bp 790bp,clip,scale=0.84]
{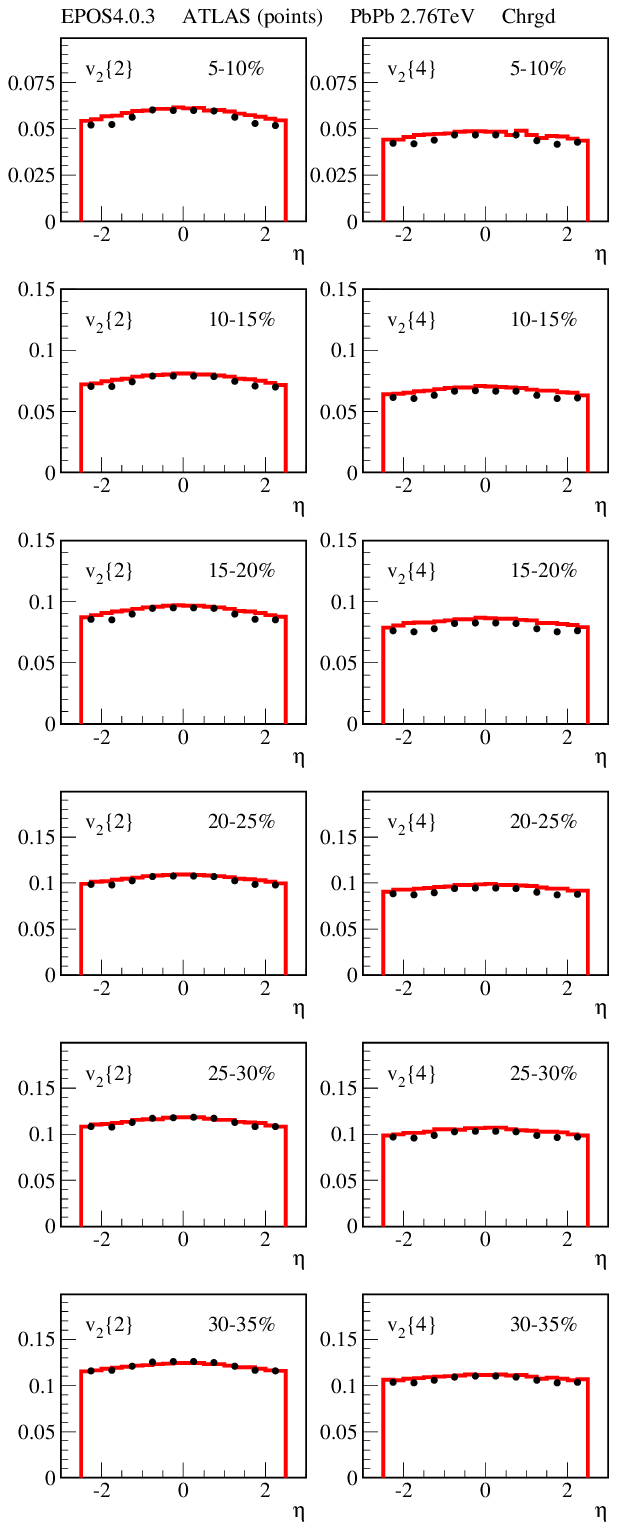}
\caption{Results for $v_{2}\{2\}$ and $v_{2}\{4\}$ versus pseudorapidity
$\eta$, in PbPb collisions at 2.76 GeV, for different centralities
(below 35\%). I compare the EPOS4 simulations using the dipole scenario
(red lines) with data from ATLAS \cite{ATLAS:2014-PbPb3-flow} (black
points). \protect\label{sel5-1}}
\end{figure}

\begin{figure}
\hspace*{-0.4cm}\includegraphics[bb=30bp 40bp 595bp 790bp,clip,scale=0.84]
{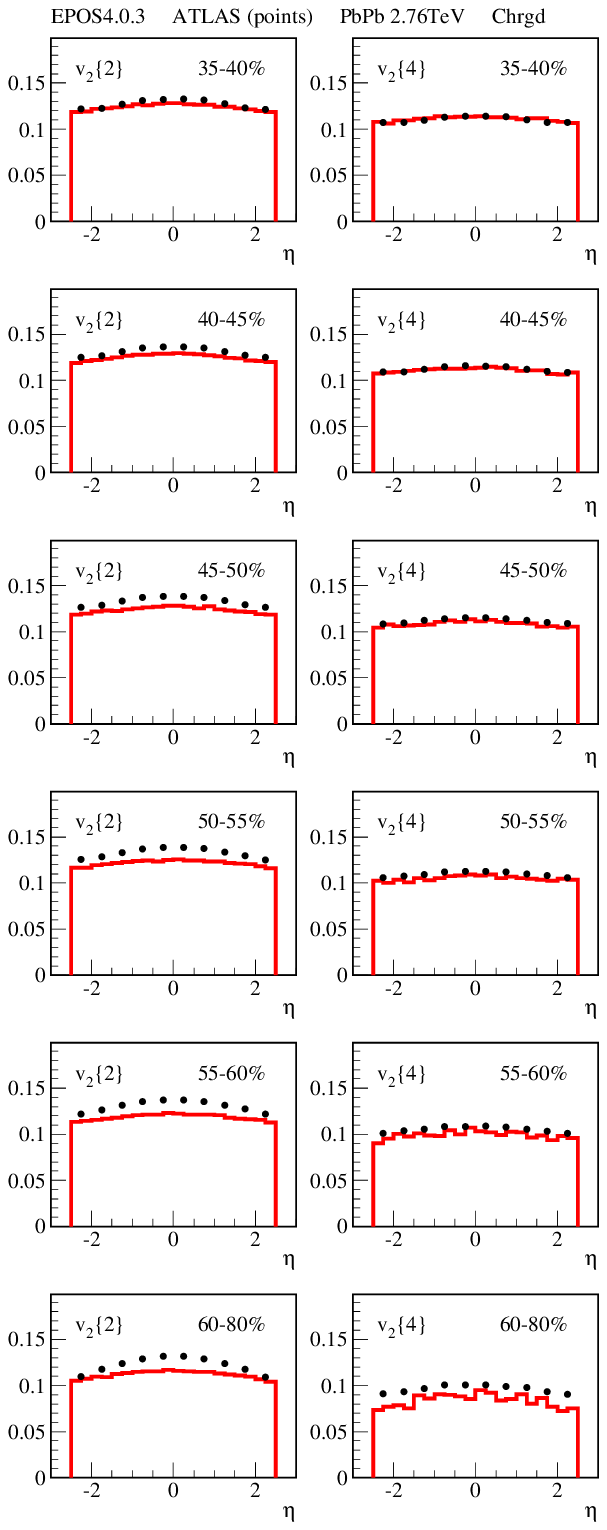}
\caption{Results for $v_{2}\{2\}$ and $v_{2}\{4\}$ versus pseudorapidity
$\eta$, in PbPb collisions at 2.76 GeV, for different centralities
(above 35\%). One compares the EPOS4 simulations using the dipole
scenario (red lines) with data from ATLAS \cite{ATLAS:2014-PbPb3-flow}
(black points). \protect\label{sel5-2}}
\end{figure}

\begin{figure}
\centering{}\hspace*{-0.7cm}\includegraphics[bb=30bp 40bp 595bp 790bp,clip,scale=0.84]
{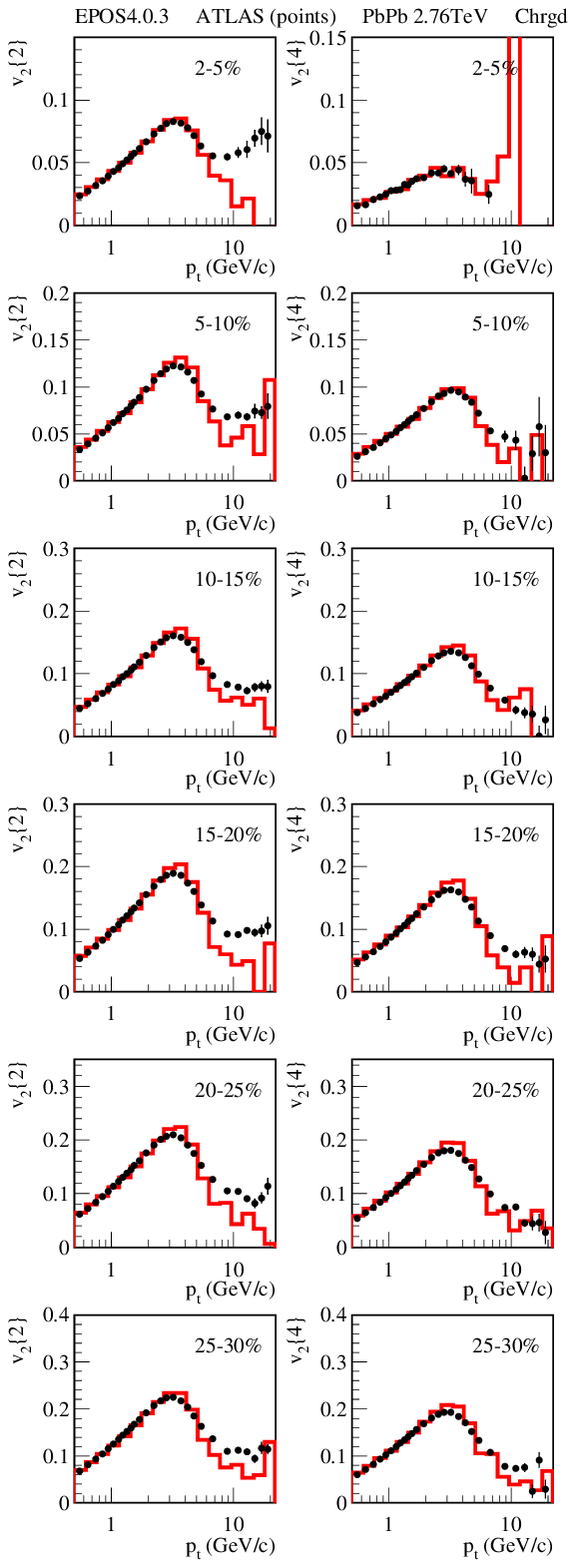}
\caption{Results for $v_{2}\{2\}$ and $v_{2}\{4\}$ versus transverse momentum,
in PbPb collisions at 2.76 GeV, for different centralities (below
30\%). One compares the EPOS4 simulations using the dipole scenario
(red lines) with data from ATLAS \cite{ATLAS:2014-PbPb3-flow} (black
points). \protect\label{sel6-1}}
\end{figure}

\begin{figure}
\centering{}\hspace*{-0.7cm}\includegraphics[bb=30bp 40bp 595bp 790bp,clip,scale=0.84]
{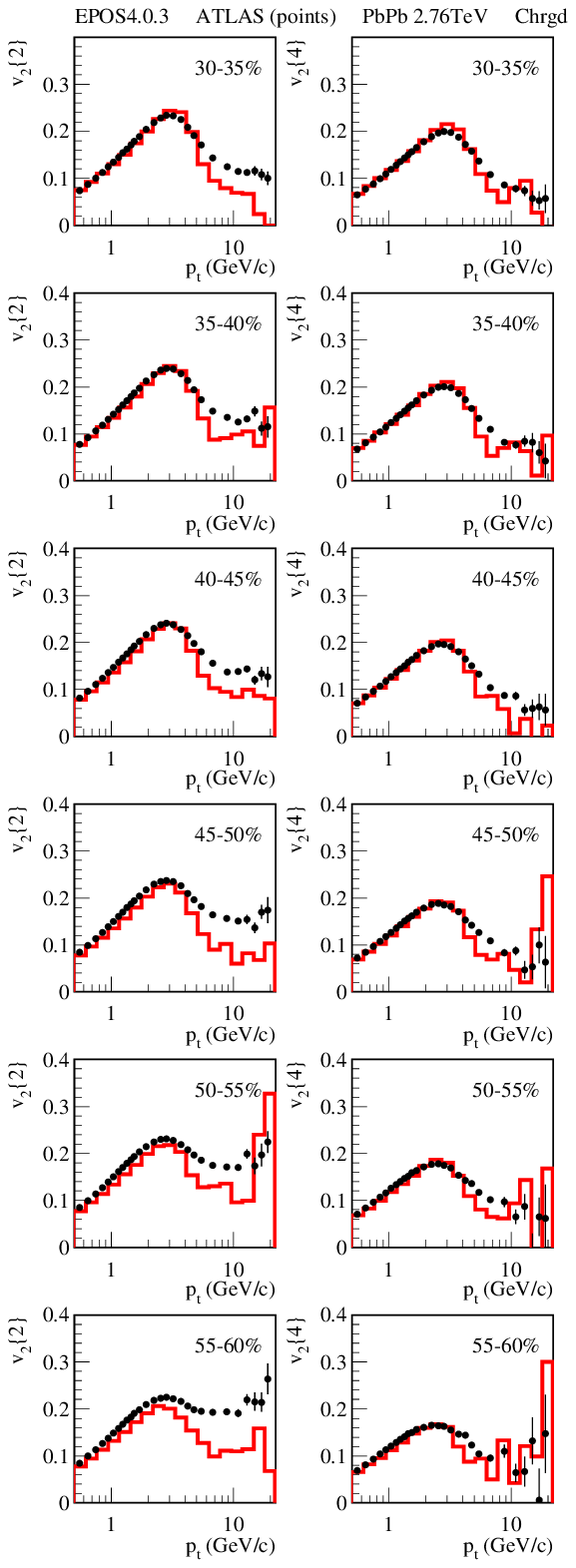}
\caption{Results for $v_{2}\{2\}$ and $v_{2}\{4\}$ versus transverse momentum,
in PbPb collisions at 2.76 GeV, for different centralities (above
30\%). One compares the EPOS4 simulations using the dipole scenario
(red lines) with data from ATLAS \cite{ATLAS:2014-PbPb3-flow} (black
points). \protect\label{sel6-2}}
\end{figure}

Whereas so far the focus was on the multiplicity (or centrality) dependence
of integrated quantities, I will present in the following (in Figs. \ref{sel5-1}, \ref{sel5-2}, \ref{sel6-1}, and \ref{sel6-2}) more detailed
results, i.e., differential quantities versus pseudorapidity or transverse
momentum, for different centrality classes, in PbPb collisions at
2.76 GeV. I do not expect new insight, but this is an exercise to
check if the approach is consistent.

In Figs. \ref{sel5-1} and \ref{sel5-2}, I show flow harmonics $v_{2}\{2\}$
and $v_{2}\{4\}$ versus pseudorapidity $\eta$, in PbPb collisions
at 2.76 GeV, for different centralities. EPOS4 simulations using the
dipole scenario (red lines) are compared with data from ATLAS \cite{ATLAS:2014-PbPb3-flow}
(black points). 

Since I have shown earlier that the PbPb results at 2.76 TeV do not
depend much on the scenario, I plot here only results from the full
EPOS4 simulations using the dipole scenario, which also represents
the default. 
In general, the simulation results are close to the data, but for
semi-peripheral collisions beyond 45\%, $v_{2}\{2\}$ is a bit low,
whereas the agreement for $v_{2}\{4\}$ is better. 
This is compatible with the centrality dependence of integrated flow
harmonics shown earlier. 


In Figs. \ref{sel6-1} and \ref{sel6-2}, I show flow harmonics $v_{2}\{2\}$
and $v_{2}\{4\}$ versus transverse momentum, in PbPb collisions at
2.76 GeV, for different centralities. EPOS4 simulations using the
dipole scenario (red lines) are compared with data from ATLAS \cite{ATLAS:2014-PbPb3-flow}
(black points). 

For low and intermediate values of $p_{t}$, the simulation results
are close to the data again, except for $v_{2}\{2\}$ in
the most peripheral events, where the simulations are slightly below
the data. 

However, the high $p_{t}$ region is not well described, simulations
are too low, indicating too little energy loss. The latter is treated
in EPOS4 in a very simplified fashion in connection with the core/corona
procedure, based on a very simple energy loss method, see Ref. \cite{werner:2023-epos4-micro}.


\section{Summary and conclusions}

I discussed in this paper the multiplicity dependence of multi-particle
cumulants and flow harmonics, to better understand collectivity in
small systems. In the first part, proton-proton collisions at 13 TeV
were investigated, where experimental findings show a flat behavior
of the elliptical flow $v_{2}\{2,|\Delta\eta|>2\}$ as a function
of the multiplicity. It has been demonstrated that the full EPOS4 simulation,
including a hydrodynamical evolution (and therefore producing flow),
cannot have such a flat behavior, if one uses the ``symmetric scenario'',
i.e., distributions of the transverse positions of the partons generated
according to a symmetric law, because here an asymmetric behavior
is entirely due to the randomness of the event-by-event generation.
The randomness disappears with an increasing number of scatterings,
associated to an increasing multiplicity. However, if one introduces
a geometric component in the form of a distribution of transverse
positions around two centers, i.e., the dipole scenario, then one
expects and observes a flat $v_{2}$ curve as in the data.

I then investigated PbPb collisions at 2.76 TeV. It was shown that
here the difference between the symmetric scenario and the dipole
scenario is small. Actually both scenarios are close to the data.
The fundamental difference between proton-proton and heavy-ion scattering
is the fact that increasing multiplicity in case of proton-proton
is due to multiple scatterings involving the same projectile and target
nucleon (there is only one), whereas in heavy-ion scattering, the
number of participating nucleons increases. In proton-proton there
is a privileged situation where both dipoles are parallel, which
is the case for all multiple scatterings, and this provides an asymmetric
initial matter distribution. In heavy-ion collisions there is a large
number of nucleon-nucleon pairs involved, and it is very unlikely
that all of them are parallel to each other, which explains that the
dipole geometry is hardly visible.

I have presented results for many different cumulants and flow harmonics
in $pp$ collisions at 13 TeV and in PbPb collisions at 2.76 TeV,
integrated and differential ones, showing that the ``dipole scenario'',
which is the default in EPOS4.0.3, gives in general a good desciption
of the data. Exceptions are the very low multiplicity region in proton-proton
collisions at 13 TeV, and the high transverse momentum region in PbPb
collisions at 2.76 TeV.

The main message of the paper, based on EPOS4 simulations compared
to data, is that a dipole form of a high-energy proton is needed to explain
flow harmonics results in small systems, and that such a scenario
gives a coherent picture when comparing simulations and a large amount
of data on multi-particle cumulants and flow harmonics. 

\medskip{}

\appendix
\noindent{\LARGE\textbf{Appendix}}{\LARGE\par}

\section{Computing flow harmonics\protect\label{_______Computing-flow-harmonics_______}}

\subsection{Cumulant method}

Following Refs. \cite{Bilandzic:2010-cumulants}%
{} and \cite{ATLAS:2014-PbPb3-flow}%
, based on \cite{Borghini:2000sa,Borghini:2001vi,Borghini:2001zr},
one first discusses the \textbf{reference flow}. For given $n$, the
\textbf{2 and 4 particle correlations} are defined as
\begin{equation}
\left\langle 2\right\rangle =\left\langle e^{in(\phi_{1}-\phi_{2})}\right\rangle =\frac{1}{w_{2}}\sum_{i,j}\,\!'\,e^{in(\phi_{i}-\phi_{j})},
\end{equation}
and
\begin{equation}
\left\langle 4\right\rangle =\left\langle e^{in(\phi_{1}+\phi_{2}-\phi_{3}-\phi_{4})}\right\rangle =\frac{1}{w_{4}}\sum_{i,j,k,l}\!\!'\,e^{in(\phi_{i}+\phi_{j}-\phi_{k}-\phi_{l})},
\end{equation}
with
\begin{align}
 & w_{2}=M(M-1),\\
 & w_{4}=M(M-1)(M-2)(M-3),
\end{align}
with $M$ being the multiplicity, counting all particles in a given
acceptance, referred to as reference flow particles (RFP), and where
$\sum'$ means that all indices in the sum must be taken different.
To compute $\left\langle 2\right\rangle $, one uses $Q_{n}$ (called
$Q$-vector, although it is actually a complex number and not a vector),
\begin{equation}
Q_{n}=\sum_{i=1}^{M}e^{in\phi_{i}}.
\end{equation}
One gets $Q_{n}\bar{Q}_{n}=M+w_{2}\left\langle 2\right\rangle $,
so
\begin{equation}
\left\langle 2\right\rangle =\frac{1}{w_{2}}\left\{ Q_{n}\bar{Q}_{n}-M\right\} \,.
\end{equation}
To compute $\left\langle 4\right\rangle $, one uses 
\begin{equation}
|Q_{n}|^{4}=Q_{n}Q_{n}\bar{Q}_{n}\bar{Q}_{n}=\sum_{i,j,k,l}e^{in(\phi_{i}+\phi_{j}-\phi_{k}-\phi_{l})}.
\end{equation}
One has four distinct cases for the indices $i$, $j$, $k$, $l$:
they are all different (4-particle correlation), three are different,
two are different or they are all the same. One finally gets
\begin{align}
 & \left\langle 4\right\rangle =\frac{1}{w_{4}}\Big\{|Q_{n}|^{4}+|Q_{2n}|^{2}-2Re\left\{ Q_{2n}\bar{Q}_{n}\bar{Q}_{n}\right\} \\
 & \qquad\qquad-4(M-2)|Q_{n}|^{2}-2M(M-3)\Big\}.\nonumber 
\end{align}
Averaging over events gives
\begin{equation}
\left\langle \left\langle 2\right\rangle \right\rangle =\sum_{events}w_{2}\left\langle 2\right\rangle /\sum_{events}w_{2},
\end{equation}
and
\begin{equation}
\left\langle \left\langle 4\right\rangle \right\rangle =\sum_{events}w_{4}\left\langle 4\right\rangle /\sum_{events}w_{4}.
\end{equation}
The \textbf{cumulants} are 
\begin{equation}
c_{n}\{2\}=\left\langle \left\langle 2\right\rangle \right\rangle ,\quad c_{n}\{4\}=\left\langle \left\langle 4\right\rangle \right\rangle -2\left\langle \left\langle 2\right\rangle \right\rangle ^{2}.
\end{equation}
This gives estimates for the \textbf{harmonics}
\begin{equation}
v_{n}\{2\}=\sqrt{c_{n}\{2\}},\;v_{n}\{4\}=\sqrt[4]{-c_{n}\{4\}}.
\end{equation}
In the following one discusses the \textbf{differential flow}, referring
to particles of interest (POI). Per definition, $M$ is the total number
of particles labeled as RFP, $m_{p}$ is the total number of particles
labeled as POI, and $m_{q}$ is the total number of particles labeled
both as RFP and POI. One defines
\begin{equation}
\left\langle 2'\right\rangle =\left\langle e^{in(\phi_{1}-\phi_{2})}\right\rangle =\frac{1}{w_{2'}}\sum_{i=1}^{m_{p}}\sum_{j=1}^{M}\!'\,e^{in(\phi_{i}-\phi_{j})},
\end{equation}
and
\begin{equation}
\left\langle 4'\right\rangle =\left\langle e^{in(\phi_{1}+\phi_{2}-\phi_{3}-\phi_{4})}\right\rangle =\frac{1}{w_{4'}}\sum_{i=1}^{m_{p}}\sum_{j,k,l=1}^{M}\!\!\!\!'\,e^{in(\phi_{i}+\phi_{j}-\phi_{k}-\phi_{l})},
\end{equation}
with
\begin{align}
 & w_{2'}=m_{p}M-m_{q},\\
 & w_{4'}=(m_{p}M-3m_{q})(M-1)(M-2),
\end{align}
where $w_{4'}$ counts the number of different indices, being the
sum of $(m_{p}-m_{q})M(M-1)(M-2)$ and $m_{q}(M-1)(M-2)(M-3)$. The
calculation of $\left\langle 2'\right\rangle $ and $\left\langle 4'\right\rangle $
will be done based on $Q_{n}$ as well as on
\begin{equation}
p_{n}=\sum_{j=1}^{m_{p}}e^{in\phi_{j}},\quad q_{n}=\sum_{k=1}^{m_{q}}e^{in\phi_{k}}.
\end{equation}
One finds
\begin{equation}
\left\langle 2'\right\rangle =\frac{1}{w_{2'}}\left\{ p_{n}\bar{Q}_{n}-m_{q}\right\} \,,
\end{equation}
and
\begin{align}
\left\langle 4'\right\rangle  & =\frac{1}{w_{4'}}\Big\{ p_{n}Q_{n}\bar{Q}_{n}\bar{Q}_{n}-q_{2n}\bar{Q}_{n}\bar{Q}_{n}\nonumber \\
 & \qquad-p_{n}Q_{n}\bar{Q}_{2n}-2Mp_{n}\bar{Q}_{n}-2m_{q}|Q_{n}|^{2}\\
 & \qquad+7q_{n}\bar{Q}_{n}-Q_{n}\bar{q}_{n}+q_{2n}\bar{Q}_{2n}\nonumber \\
 & \qquad+2p_{n}\bar{Q}_{n}+2m_{q}M-6m_{q}\Big\}.\nonumber 
\end{align}
Averaging over events gives
\begin{equation}
\left\langle \left\langle 2'\right\rangle \right\rangle =\sum_{events}w_{2'}\left\langle 2'\right\rangle /\sum_{events}w_{2'}\;,
\end{equation}
\begin{equation}
\left\langle \left\langle 4'\right\rangle \right\rangle =\sum_{events}w_{4'}\left\langle 4'\right\rangle /\sum_{events}w_{4'}\;.
\end{equation}
The \textbf{cumulants} are 
\begin{equation}
c'_{n}\{2\}=\left\langle \left\langle 2'\right\rangle \right\rangle ,\quad c'_{n}\{4\}=\left\langle \left\langle 4'\right\rangle \right\rangle -2\left\langle \left\langle 2'\right\rangle \right\rangle \left\langle \left\langle 2\right\rangle \right\rangle .
\end{equation}
This gives estimates for the \textbf{harmonics}
\begin{equation}
v'_{n}\{2\}=\frac{c'_{n}\{2\}}{\sqrt{c_{n}\{2\}}},\;v'_{n}\{4\}=\frac{-c'_{n}\{4\}}{\sqrt[4]{-c_{n}\{4\}}^{3}}.
\end{equation}

One may introduce a pseudorapidity gap $\Delta\eta$ for two-particle correlations, by considering for the reference flow two sets of particles in distinct $\eta$ intervals $a$ and $b$ with a gap  $\Delta\eta$ between them. This requires a minor modification of the above formulas.  Instead of a multiplicity $M$, one has two multiplicities $M_a$ and $M_b$, and for the weight $w_2$  one has $w_2=M_a M_b$. Instead of $Q_n$ one has two $Q$-vectors $Q_n^a$ and $Q_n^b$, and instead of $\{{Q_n \bar{Q}_n}-M\}$ one takes the real part of   $Q_n^a \bar{Q}_n^b$. 

\subsection{Scalar product method}

Following the Refs. \cite{STAR:2002-Elliptic,ALICE:2014-Elliptical},%
one defines three sets of particles, in different pseudorapidity ranges,
for example:
\begin{description}
\item [{A}] Particles in the range $-3.7<\eta<-1.7$ ~~~($N_{A}$ reference
particles)
\item [{B}] Particles in the range $-1<\eta<1$ ~~~($N_{B}$ labeled
particles)
\item [{C}] Particles in the range $2.8<\eta<5.1$ ~~~($N_{C}$ reference
particles)
\end{description}
Then one defines a two-dimensional vector $\vec{Q}_{n}^{S}$ for $S$
being any of the sets A or C as

\begin{equation}
\vec{Q}_{n}^{S}=\left(\sum_{S}\mathrm{cos}(n\varphi)\,,\,\sum_{S}\mathrm{sin}(n\varphi)\right),
\end{equation}
and $\left.\vec{q}\right._{n}^{S}=\vec{Q}_{n}^{S}/N_{S}$, so 
\begin{equation}
\left.\vec{q}\right._{n}^{S}=\left(\left\langle \mathrm{cos}(n\varphi)\right\rangle _{S}\,,\,\left\langle \mathrm{sin}(n\varphi)\right\rangle _{S}\right).
\end{equation}
For each particle in B, one defines 
\begin{equation}
\vec{u}_{n}^{B}=\left(\mathrm{cos}(n\varphi)\,,\,\mathrm{sin}(n\varphi)\right).
\end{equation}
The elliptical flow is then
\[
v_{2}=\sqrt{\frac{\left\langle \left\langle \vec{u}_{2}^{B}\cdot\vec{q}_{2}^{A}\right\rangle \right\rangle \left\langle \left\langle \vec{u}_{2}^{B}\cdot\vec{q}_{2}^{C}\right\rangle \right\rangle }{\left\langle \vec{q}_{2}^{A}\cdot\vec{q}_{2}^{C}\right\rangle }}.
\]

\section{High multiplicity events \protect\label{_______High=000020multiplicity=000020events_______-1}}

\begin{figure}[h]
\centering{}\hspace*{-0.2cm}\includegraphics[bb=17bp 30bp 595bp 530bp,clip,scale=0.5]
{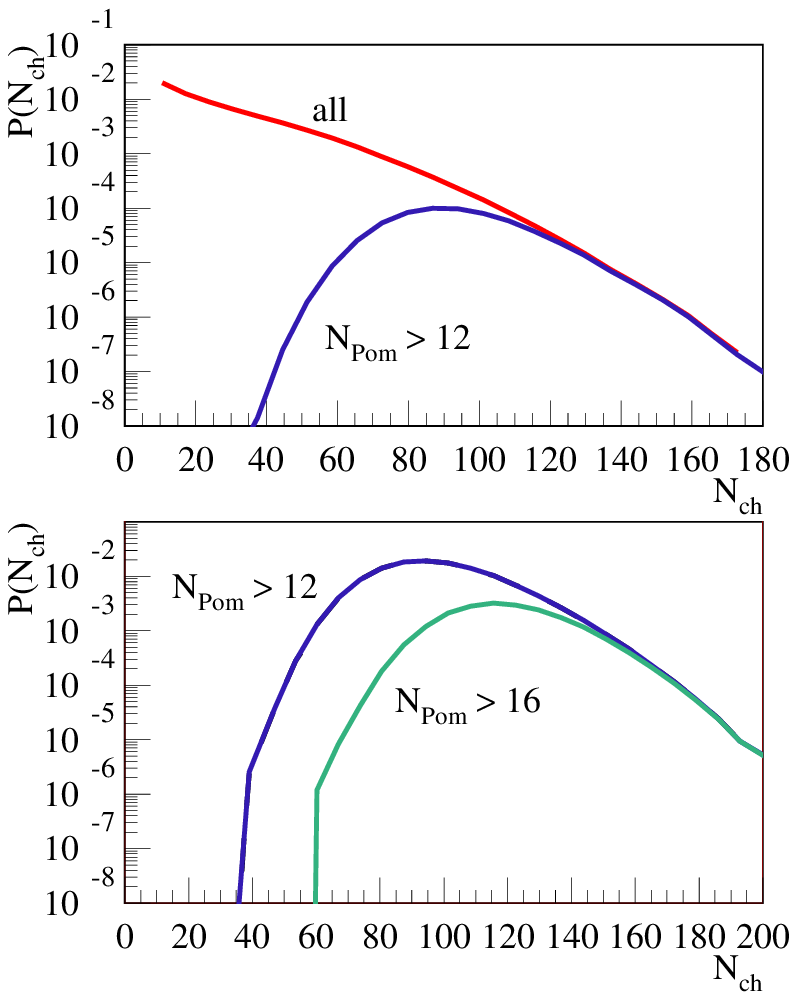}
\caption{Multiplicity ($N_{\mathrm{ch}}$) distributions in pp scattering at
13 TeV, where ``all'' refers to minimum bias results, ``$N_{\mathrm{Pom}}>12$''
refers to events with more than 12 Pomerons, and ``$N_{\mathrm{Pom}}>16$''
refers to events with more than 16 Pomerons. \protect\label{sel15}}
\end{figure}

In Fig. \ref{sel15}, multiplicity distributions in pp scattering
at 13 TeV are shown, where ``all'' refers to minimum bias results,
``$N_{\mathrm{Pom}}>12$'' refers to events with more than 12 Pomerons,
and ``$N_{\mathrm{Pom}}>16$'' refers to events with more than 16
Pomerons.
This shows that the number of Pomerons can be used as a trigger to
generate high multiplicity events.\bigskip{}

\end{document}